\documentclass[useAMS,usenatbib]{mn2e}
\usepackage[dvips]{color}
\usepackage{amssymb,graphicx,multirow,txfonts}

\newcommand{\etal}{et~al.}
\newcommand{\PVdblt}{{\rm P}\kern 0.1em{\sc v}~$\lambda\lambda 1117, 1128$}
\newcommand{\CaIIdblt}{{\rm Ca}\kern 0.1em{\sc ii}~$\lambda\lambda 3934, 3969$}
\newcommand{\AlIIIdblt}{{\rm Al}\kern 0.1em{\sc iv}~$\lambda\lambda 1855, 1863$}
\newcommand{\CIVdblt}{{\rm C}\kern 0.1em{\sc iv}~$\lambda\lambda 1548, 1550$}
\newcommand{\MgIIdblt}{{\rm Mg}\kern 0.1em{\sc ii}~$\lambda\lambda 2796, 2803$}
\newcommand{\NVdblt}{{\rm N}\kern 0.1em{\sc v}~$\lambda\lambda 1238, 1242$}  
\newcommand{\SVIdblt}{{\rm S}\kern 0.1em{\sc vi}~$\lambda\lambda 933, 944$} 
\newcommand{\OVIdblt}{{\rm O}\kern 0.1em{\sc vi}~$\lambda\lambda 1031, 1037$} 
\newcommand{\SiIIdblt}{{\rm Si}\kern 0.1em{\sc ii}~$\lambda\lambda 1190, 1193$} 
\newcommand{\SiIVdblt}{{\rm Si}\kern 0.1em{\sc iv}~$\lambda\lambda 1393, 1402$} 
\newcommand{\AlI}{\hbox{{\rm Al}\kern 0.1em{\sc i}}}
\newcommand{\AlII}{\hbox{{\rm Al}\kern 0.1em{\sc ii}}}
\newcommand{\AlIII}{{\hbox{\rm Al}\kern 0.1em{\sc iii}}}
\newcommand{\CaII}{\hbox{{\rm Ca}\kern 0.1em{\sc ii}}}
\newcommand{\CII}{\hbox{{\rm C}\kern 0.1em{\sc ii}}}
\newcommand{\CIIe}{\hbox{{\rm C$^{\ast}$}\kern 0.1em{\sc ii}}}
\newcommand{\CIII}{\hbox{{\rm C}\kern 0.1em{\sc iii}}}
\newcommand{\CIV}{\hbox{{\rm C}\kern 0.1em{\sc iv}}}
\newcommand{\CV}{\hbox{{\rm C}\kern 0.1em{\sc v}}}
\newcommand{\HI}{\hbox{{\rm H}\kern 0.1em{\sc i}}}
\newcommand{\HII}{\hbox{{\rm H}\kern 0.1em{\sc ii}}}
\newcommand{\Lya}{\hbox{{\rm Ly}\kern 0.1em$\alpha$}}
\newcommand{\Lyb}{\hbox{{\rm Ly}\kern 0.1em$\beta$}}
\newcommand{\Lyg}{\hbox{{\rm Ly}\kern 0.1em$\gamma$}}
\newcommand{\Lyd}{\hbox{{\rm Ly}\kern 0.1em$\delta$}}
\newcommand{\HeI}{\hbox{{\rm He}\kern 0.1em{\sc i}}}
\newcommand{\HeII}{\hbox{{\rm He}\kern 0.1em{\sc ii}}}
\newcommand{\FeI}{\hbox{{\rm Fe}\kern 0.1em{\sc i}}}
\newcommand{\FeII}{\hbox{{\rm Fe}\kern 0.1em{\sc ii}}}
\newcommand{\FeIII}{\hbox{{\rm Fe}\kern 0.1em{\sc iii}}}
\newcommand{\MnII}{\hbox{{\rm Mn}\kern 0.1em{\sc ii}}}
\newcommand{\MgI}{\hbox{{\rm Mg}\kern 0.1em{\sc i}}}
\newcommand{\MgII}{\hbox{{\rm Mg}\kern 0.1em{\sc ii}}}
\newcommand{\MgIII}{\hbox{{\rm Mg}\kern 0.1em{\sc iii}}}
\newcommand{\NI}{\hbox{{\rm N}\kern 0.1em{\sc i}}}
\newcommand{\NII}{\hbox{{\rm N}\kern 0.1em{\sc ii}}}
\newcommand{\NIII}{\hbox{{\rm N}\kern 0.1em{\sc iii}}}
\newcommand{\NV}{\hbox{{\rm N}\kern 0.1em{\sc v}}}
\newcommand{\OVI}{\hbox{{\rm O}\kern 0.1em{\sc vi}}}
\newcommand{\OI}{\hbox{{\rm O}\kern 0.1em{\sc i}}}
\newcommand{\OII}{\hbox{[{\rm O}\kern 0.1em{\sc ii}]}}
\newcommand{\OIII}{\hbox{[{\rm O}\kern 0.1em{\sc iii}]}}
\newcommand{\OIV}{\hbox{{\rm O}\kern 0.1em{\sc iv}]}}
\newcommand{\SI}{{\rm S}\kern 0.1em{\sc i}}
\newcommand{\SIV}{{\rm S}\kern 0.1em{\sc iv}}
\newcommand{\SVI}{{\rm S}\kern 0.1em{\sc vi}}
\newcommand{\SiI}{\hbox{{\rm Si}\kern 0.1em{\sc i}}}
\newcommand{\SiII}{\hbox{{\rm Si}\kern 0.1em{\sc ii}}}
\newcommand{\SiIII}{\hbox{{\rm Si}\kern 0.1em{\sc iii}}}
\newcommand{\SiIV}{\hbox{{\rm Si}\kern 0.1em{\sc iv}}}
\newcommand{\SII}{\hbox{{\rm S}\kern 0.1em{\sc ii}}}
\newcommand{\SIII}{\hbox{{\rm S}\kern 0.1em{\sc iii}}}
\newcommand{\NaI}{\hbox{{\rm Na}\kern 0.1em{\sc i}}}
\newcommand{\TiII}{\hbox{{\rm Ti}\kern 0.1em{\sc ii}}}
\newcommand{\ZnII}{\hbox{{\rm Zn}\kern 0.1em{\sc ii}}}
\newcommand{\CrII}{\hbox{{\rm Cr}\kern 0.1em{\sc ii}}}
\newcommand{\kms}{\hbox{km~s$^{-1}$}}
\newcommand{\cmsq}{\hbox{cm$^{-2}$}}

\def\aj{{AJ}}                   
\def\araa{{ARA\&A}}             
\def\apj{{ApJ}}                 
\def\apjl{{ApJ}}                
\def\apjs{{ApJS}}

\def\aap{{A\&A}}

\def\mnras{{MNRAS}}

\def\jcp{{J.~Chem.~Phys.}}

\title[Morphological properties of $z\sim 0.5$ absorbers]{Morphological properties of $\bmath{z\sim0.5}$ absorption-selected galaxies: the role of galaxy inclination}

\author[G. G. Kacprzak et al.]{Glenn G. Kacprzak,$^{1}$\thanks{gkacprzak@astro.swin.edu.au} 
Christopher W. Churchill,$^{2}$
Jessica L. Evans,$^{2}$
\newauthor Michael T. Murphy,$^{1}$ and 
Charles C. Steidel$^{3}$\\
$^{1}$ Centre for Astrophysics and Supercomputing, Swinburne University of Technology, PO Box 218, Victoria 3122, Australia\\
$^{2}$ Department of Astronomy, New Mexico State University, Las Cruces, NM 88003\\
$^{3}$ California Institute of Technology, MS 105-24, Pasadena, CA 91125, USA}

\begin{document}
\date{Accepted June 15, 2011}

\pagerange{3118--3137} \pubyear{2011}

\maketitle

\label{firstpage}

\begin{abstract}
\noindent We have used GIM2D to quantify the morphological properties
of 40 intermediate redshift {\MgII} absorption-selected galaxies
($0.03\leq W_r(2796) \leq 2.9$~\AA), imaged with WFPC-2/{\it HST}, and
compared them to the halo gas properties measured from HIRES/Keck and
UVES/VLT quasar spectra.  We find that as the quasar--galaxy
separation, $D$, increases the {\MgII} equivalent decreases with large
scatter, implying that $D$ is not the only physical parameter
affecting the distribution and quantity of halo gas. Our main result
shows that inclination correlates with {\MgII} absorption properties
after normalizing out the relationship (and scatter) between the
absorption properties and $D$. We find a $4.3\sigma$ correlation
between $W_r(2796)$ and galaxy inclination, normalized by impact
parameter, $i/D$.  Other measures of absorption optical depth also
correlate with $i/D$ at greater than $3.2\sigma$ significance.
Overall, this result suggests that {\MgII} gas has a co-planer
geometry, not necessarily disk-like, that is coupled to the galaxy
inclination.  It is plausible that the absorbing gas arises from tidal
streams, satellites, filaments, etc., which tend to have somewhat
co-planer distributions.  This result does not support a picture in
which {\MgII} absorbers with $W_r(2796)\la 1$\,\AA\ are predominantly
produced by star-formation driven winds.

We further find that; (1) {\MgII} host galaxies have quantitatively
similar bulge and disk scale length distribution to field galaxies at
similar redshifts and have a mean disk and bulge scale length of
3.8~kpc and 2.5~kpc, respectively; (2) Galaxy color and luminosity do
not correlate strongly with absorption properties, implying a lack of
a connection between host galaxy star formation rates and absorption
strength; (3) Parameters such as scale lengths and bulge-to-total
ratios do not significantly correlate with the absorption parameters,
suggesting that the absorption is independent of galaxy size or mass.
   
\end{abstract}

\begin{keywords}
---galaxies: ISM, haloes  ---quasars: absorption lines.
\end{keywords}

\section{Introduction}

Since the first observational evidence associating foreground galaxies
with absorption lines detected in the spectra of background quasars
\citep{boksenberg78}, researchers have strived to determine the
relationships between absorbing gas found within $\sim$200~{\kms} and
a few 100~kpc of their host galaxies.  We have yet to develop a
complete understanding of the many physical conditions under which
galaxies predominantly produce their own metal--enriched ``halos'' or
accrete material in their immediate environment.  It is plausible that
most galaxies undergo both these processes at some point during their
evolution.  Assuming that clear observational connections between the
galaxy properties and the halo gas properties are present in nature,
our collective hope is that we can ultimately understand the
conditions that drive outflows versus the conditions that indicate
accretion (such as galaxy star formation rates, morphologies,
inclinations, etc., and absorption equivalent widths, kinematics,
and chemical and ionization conditions).

The {\MgIIdblt} doublet is commonly used to trace metal-enriched
low-ionized gas surrounding galaxies between $0.1\leq z \leq 2.5$.
However, understanding the origins of this gas is difficult.  {\MgII}
absorption traces a wide range of neutral hydrogen column densities,
$10^{16} \lesssim \hbox{N(\HI)} \lesssim 10^{22}$~{\cmsq}
\citep{archiveI,weakII}, which produces a large range of detectable
{\MgII} rest-frame equivalent widths, $0.02 \lesssim W_r(2796)\lesssim
10$~\AA.  The large range in {\HI} column densities implies that
{\MgII} absorption arises within a large dynamical range of structures
and environments that contribute to the complex kinematics seen in
typical absorption profiles.  It may be that lower column density
systems trace a different population of structures and gas producing
processes than do higher column density systems.

The first suggestion that halos observed via {\MgII} absorption may
exhibit a dependence on galaxy--quasar separation was a $\sim 3\sigma$
anti-correlation between $W_r(2796)$ and impact parameter, $D$
\citep[e.g.,][]{lanzetta90,steidel95,archiveII}.  This could be
interpreted to imply the universal property of a radially decreasing
gas density profile in halos.  However, given the complex velocity
structure of these absorption systems, which is independent of $D$,
\citet{csv96} argued that the absorption likely arises from a variety
of ongoing dynamical events within the galaxy and halo.

The first hint of a connection between galaxy mass and the presence or
absence of absorption was deduced from the Holmberg-like luminosity
scaling between a characteristic halo radius and galaxy K-band
luminosity \citep{steidel95}.  Recent studies at $0.2 \leq z \leq 1$
restrict the range of the luminosity power-law slope between 0.2
$\lesssim \beta \lesssim$ 0.5, the characteristic {\MgII} halo radius
between 90--110~kpc (for an $L_B^{\star}$ galaxy), and the gas
absorption covering fraction between $\sim$50--90\%
\citep{tripp-china,chen08,kacprzak08,chen10}.  However, the {\MgII}
gas covering fraction may decrease to $\sim$25\% by $z=0.1$
\citep{barton09}.  It remains unclear what physical processes
replenish the gas reservoirs and populate halos with such high
covering fractions at high impact parameters.

\citet{bond01b} analyzed several systems with $W_r(2796)>1.8$~{\AA} at
$z>1$ and concluded that high equivalent width absorption systems
produced by winds is a plausible model.  \citet{bouche06} found an
anti-correlation between $W_r(2796)$ and the amplitude of the
cross-correlation between luminous red galaxies (LRGs) and {\MgII}
absorbers with $W_r(2796)>1$~\AA.  Since the cross-correlation
amplitude is related to the halo mass, they inferred that the majority
of strong absorption systems are {\it not\/} produced by gas that is
virialized within galaxy halos. They interpreted the
$W_r(2796)$--halo-mass anti-correlation as evidence that
supernova-driven winds become dominant sources of stronger {\MgII}
absorbers, i.e., $W_r(2796)>2$~{\AA}.  Further support for a wind
interpretation of such strong absorbers was found by \citet{zibetti07}
by stacking Sloan g, r, and i band images toward 700 quasars.
Spectral energy distribution fits to their stacked images indicate
that early-type galaxies are associated with weaker absorption than
late-type galaxies, which suggests that star formation rates (SFRs)
may correlate with {\MgII} absorption strength.  Confirming these
results, \citet{menard09} stacked over 8,500 segments of SDSS quasar
spectra where {\OII} was likely to be present at the redshift of the
{\MgII} absorption and discovered a highly significant correlation
between the $W_r(2796)$ and {\OII} luminosity.  They further
demonstrated that the distribution function of $W_r(2796)$ naturally
reproduces the shape and amplitude of the {\OII} luminosity without
any free parameters.  This implies that {\MgII} absorption systems
trace a significant fraction of the global {\OII} luminosity
function. They suggest that star-formation driven outflows are the
primary mechanism responsible for $W_r(2796)\gtrsim 1$~{\AA}
absorption systems.

The outflow scenario is further supported by the 500--2000~{\kms}
winds observed in the high equivalent width and asymmetric {\MgII}
absorption profiles seen in spectra of highly star-forming galaxies
\citep{tremonti07,weiner09,martin09,rubin09}. \citet{weiner09} further
demonstrated that $W_r(2796)$ and outflow velocity correlate with
galaxy SFR.

The evidences discussed above for strong {\MgII} systems being
produced by star-formation driven winds primarily uses data that have
high detection limits such that $W_r(2796)\gtrsim0.5$~{\AA} and it is
likely that these conclusions apply to systems with $W_r(2796) >
1$$-$$2$~{\AA}.  In a sample of 71 absorbing and non-absorbing
galaxies consisting of predominantly (85\%) $W_r(2796) < 1$~{\AA}
absorption systems, \citet{chen10} do not find a correlation between
$W_r(2796)$ and galaxy color and suggest a lack of a physical
connection between the {\MgII} halos and recent star formation history
of the galaxies. Furthermore, \citet{chen10b} found that the {\MgII}
halo size scales with the stellar mass and the specific star formation
rate of the host galaxy.  They interpret this result as massive
galaxies having more extended halos and that {\MgII} absorption
systems arise from infalling clouds that fuel star formation. Taken
together, the \citet{bond01b}, \citet{bouche06}, \citet{zibetti07},
\citet{menard09}, and \citet{chen10, chen10b} results may be
suggesting that $W_r(2796) < 1$-$1.5$~{\AA} absorption systems are not
wind dominated but that larger $W_r(2796)$ systems are wind dominated.
Studies that incorporate the gas kinematics, galaxy morphologies and
galaxy orientations relative to the quasar line of sight over the full
range of $W_r(2796)$ would be useful to examine these ideas.

\citet{kacprzak10a} compared {\MgII} absorption and galaxy rotation
kinematics of 10, $z\sim 0.5$, $W_r(2796) < 1.4$~{\AA} systems and
found that, in most cases, the absorption was fully to one side of the
galaxy systemic velocity and usually aligned with one arm of the
rotation curve. These results are consistent with an earlier study of
six galaxies by \citet{steidel02}.  However, both studies demonstrated
that in virtually all cases, a co-rotating halo model poorly
reproduces the {\MgII} absorption velocity spread. This implies that
although disk rotation may account for some of the halo gas
kinematics, other mechanisms must be invoked to account for the full
velocity spreads.  In only 3/17 cases studied in this manner, all of
which had $W_r(2796)>1$~{\AA}, the absorption kinematics of the
systems displayed possible signatures of winds or superbubbles
\citep{bond01a,ellison03,kacprzak10a}.  Using quasar absorption line
methods through cosmological simulations, \citet{kacprzak10a}
demonstrated that the majority of the {\MgII} absorption arose in an
array of structures, such as filaments and tidal streams, which were
all infalling toward the galaxy with velocities in the range of the
rotation velocity of the simulated galaxy.

\citet{kacprzak11} also directly compared the relative {\MgII} halo
gas and host galaxy kinematics for 13~ $\sim$$L_{\star}$ galaxies at
$z\sim 0.1$.  They found that these galaxies had low SFRs, low SFRs
per unit area, and their ${\hbox{{\rm Na}\kern 0.1em{\sc i}}{\rm D}}$
(stellar+ISM tracer) and {\hbox{{\rm Mg}\kern 0.1em{\sc i}}{\rm b}}
(stellar tracer) line ratios implied kinematically quiescent
interstellar media containing no strong outflowing gas. Given that
these host galaxies were in isolated environments and given the
relative halo-gas/galaxy velocity offsets, \citet{kacprzak11}
suggested a scenario in which the cool halo gas was infalling and
providing a gas reservoir that could maintain the low levels of star
formation within the host galaxies.

The majority of these studies do not incorporate quantitative
morphological and geometric parameters of the host galaxies.  Using
ground-based imaging studies of {\MgII} absorbing galaxies,
\citet{sdp94} determined that although most galaxy colors are
represented, the absorbing galaxies have an average $B - K$ color
consistent with that of a local Sb galaxy \citep[also
see][]{zibetti07,chen10}.  To date, high resolution space-based
WFPC-2/{\it HST} images of {\MgII} absorption-selected galaxies have
been used only to qualitatively state that their morphological types
appear similar to those of local spiral and elliptical galaxies
\citep{steidel98,chen01,chen03}.

In a first effort to quantify morphologies of {\MgII} absorbers,
\citet{kacprzak07} used a two-dimensional decomposition fitting
program GIM2D \citep{simard02} to model the physical structural
parameters of galaxies and compared them to {\MgII} absorption
properties.  They found a correlation between the galaxy morphological
asymmetry, normalized by impact parameter, and $W_r(2796)$. Their
correlation increases in significance when strong systems
[$W_r(2796)>1.4$\,\AA] were removed (the greater scatter for stronger
systems perhaps suggesting a different mechanism giving rise to the
gas in stronger systems). Furthermore, they reported that {\MgII}
absorption-selected galaxies have higher levels of morphological
perturbations than found for typical field galaxies.  Their
correlation suggests that known processes responsible for populating
halos with gas, such as satellite mergers and longer range
galaxy--galaxy interactions, also induce minor perturbations observed
in the morphologies of the host galaxies.

The ability to use the WFPC-2/{\it HST} images for more than measuring
magnitudes or impact parameters, by actually quantifying the
properties of the galaxies, is a useful way of exploring the
absorber--galaxy connection and may shed light on the differences in
results found for different equivalent width regimes.  In an effort to
understand the causal connection between absorbing halo gas and its
host galaxy(ies), we expand on our previous work of \citet{kacprzak07}
and apply GIM2D modeling to WFPC-2/{\it HST} images to quantify
additional morphological properties of known {\MgII} absorbing
galaxies.  In \citet{kacprzak07} we focused on morphological
asymmetries of absorption-selected galaxies and how they differ from
an ``ideal'' galaxy, which was modeled by exponential disk and a de
Vaucouleurs bulge ($n=4$). In this paper, we fit a more realistic
model to 40 galaxies, 34 from \citet{kacprzak07} plus an additional 6
galaxies selected from the literature, using an exponential disk and a
S{\'e}rsic profile bulge ($0.2\leq n\leq 4.0$) in order to quantify an
additional twelve galaxy morphological parameters.  We also compare
the galaxy morphological properties to the associated {\MgII}
absorption properties (including kinematics) and explore plausible
relationships between the absorbing gas and host galaxy properties In
\S~\ref{sec:data}, we describe our sample and analysis. In
\S~\ref{sec:results}, we present the measured properties of our sample
and identify correlations between galaxy and absorption properties. In
\S~\ref{sec:dis}, we discuss what can be inferred from the results.
Concluding remarks are offered in \S~\ref{sec:conclusion}. Throughout
we adopt an H$_{\rm 0}=70$~\kms Mpc$^{-1}$, $\Omega_{\rm M}=0.3$,
$\Omega_{\Lambda}=0.7$ cosmology.

\section{GALAXY SAMPLE AND DATA ANALYSIS}
\label{sec:data}

We have constructed a sample of 40 galaxies, with spectroscopically
confirmed redshifts 0.3$<$$z$$<$1.0, selected by the presence of
{\MgII} absorption in quasar spectra ($W_r(2796)\geq0.03$~\AA).  We
selected 34 galaxies from \citet{kacprzak07} and 6 additional galaxies
found in the literature\footnote{Three galaxies are not used from the
\citet{kacprzak07} sample: Q1127-145G1 and G2 with $z_{abs}=0.312710$
were discovered to be in a group of at least five galaxies at similar
redshifts close to the quasar line-of-sight
\citep{kacprzak10b}. Q2206-199G1 with $z_{abs}=0.751923$ is likely a
star \citep{kacprzak10a}.}. The absorption properties were measured
from HIRES/Keck \citep{vogt94} and UVES/VLT \citep{dekker00}
spectra. Galaxy properties were measured and modeled from F702W or
F814W WFPC-2/{\it HST} images of the quasar
fields. Figure~\ref{fig:mos} summarizes the absorption selection,
spectroscopic analysis and morphological fitting technique. Note that
three absorbers have two galaxies at the same redshifts
(Q$0450$$-$$132$, Q$1127$$-$$145$ and Q$1623$$+$$268$). Since it is
possible that a pair of galaxies can give rise to the absorption and
such conditions can provide further constraints into the
galaxy--absorption connection, we do not exclude these pairs from our
sample. Below we describe our data and modeling techniques.

\subsection{Quasar Spectroscopy}

\begin{table}
\begin{center}
  \caption{Keck/HIRES and VLT/UVES high resolution quasar
observations. The table columns are (1) the quasar field, (2) the
quasar redshift, (3) the instrument used, (4) the observation date(s),
and (5) the integration time in seconds.}
  \vspace{-0.5em}
\label{tab:qsospec}
{\footnotesize\begin{tabular}{lcccr}\hline
             & $z_{em}$ & Instru- &   & Exposure\\
QSO Field       &    &ment&Date (UT) &(seconds)\\\hline
Q$0002$$+$$051$ &1.90&HIRES & 1994 Jul. 05  & 2700\\
Q$0117$$+$$213$ &1.49&HIRES & 1995 Jan. 23  & 5400\\
Q$0229$$+$$131$ &2.06&HIRES & 1999 Feb. 08  & 3600\\
Q$0349$$-$$146$ &0.62&UVES  &  $\phantom{0}$2005 Oct. 11$^a$ & 1200\\
Q$0450$$-$$132$ &2.25&HIRES & 1995 Jan. 24  & 5400\\
Q$0454$$-$$220$ &0.53&HIRES & 1995 Jan. 22  & 5400\\
Q$0827$$+$$243$ &0.94&HIRES & 1998 Feb. 27  & 7200\\
Q$0836$$+$$113$ &2.70&HIRES & 1998 Feb. 26  & 5400\\
Q$1038$$+$$064$ &1.27&HIRES & 1998 Mar. 01  & 7200\\
Q$1127$$-$$145$ &1.18&UVES  & $ \cdots$$^a$  & 24,900\\
Q$1148$$+$$387$ &1.30&HIRES & 1995 Jan. 24  &5400\\  
Q$1209$$+$$107$ &2.19&UVES  &  $\phantom{0}$2003 Mar. 12$^a$  & 14,400\\
Q$1222$$+$$228$ &2.05&HIRES & 1995 Jan. 22  &3600\\
Q$1241$$+$$176$ &1.27&HIRES & 1995 Jan. 22  &2400\\
Q$1246$$-$$057$ &2.24&HIRES & 1998 Mar. 01  & 3600\\
Q$1317$$+$$277$ &1.02&HIRES & 1995 Jan. 23  &3600\\
Q$1354$$+$$195$ &0.72&HIRES & 1995 Jan. 22  &3600\\
Q$1424$$-$$118$ &0.81&UVES  & 2005 Jul. 30  & 1440 \\
Q$1622$$+$$235$ &0.93&UVES  &  $\phantom{0}$2003 Jul. 18$^a$  & 9800\\

Q$1623$$+$$268$ &2.52&HIRES & $\phantom{0}$1995 Aug. 20$^b$  & 50,360\\
Q$2128$$-$$123$ &0.50&HIRES & 1995 Aug. 20 & 2500\\
Q$2206$$-$$199$ &2.56&UVES  & $ \cdots$$^a$  & 53,503\\\hline
\end{tabular}}
\end{center}
$^a$ The PID for Q$0349-146$ is 076.A-0860(A). The Q$1127-145$ quasar
spectrum was obtained over multiple nights. The PIDs for this quasar
are 67.A-0567(A) and 69.A-0371(A). The PID for Q$1209$+$107$ is
68.A-0170(A). The PID for Q$1424-118$ is 075.A-0841(A). The PID for
Q$1622+235$ is 69.A-0371(A). The Q$2206-199$ quasar spectrum was also
obtained over multiple nights for PIDs 65.O-0158(A), 072.A-0346(A),
and 074.A-0201(A).  

$^b$ The Q$1623$$+$$268$ spectrum was obtained over multiple nights;
1995 Aug. 21, 1995 Sept. 10, 1996 May 25, 2005 May 31, 2005 June 01,
2006 Jul. 02, and  2008 Sept. 25.

\end{table}

\begin{table*}
\begin{center}
  \caption{{\it HST\/}/WFPC-2 Quasar Field Observations. The table
columns are (1) the quasar field, (2) the quasar redshift, (2) the
quasar right ascension and declination (3), (4) the WFPC-2 filter, (5)
the exposure time in seconds, and (6) the proposal identification
number and primary investigator of the WFPC-2 observations.}
  \vspace{-0.5em}
\label{tab:qsoHST}
{\footnotesize\begin{tabular}{lcccrl}\hline
                & QSO -- RA    & QSO -- DEC       &       & Exposure & \\
QSO Field       &(J2000)       &(J2000)           &Filter &(seconds)  &PID/PI \\\hline
Q$0002$$+$$051$ &00:05:20.216  & $ +$05:24:10.80  & F702W & 4600 &  5984/Steidel \\
Q$0109$$+$$200$ &01:12:10.187  & $ +$20:20:21.79  & F702W & 1800 &  6303/Disney   \\
Q$0117$$+$$213$ &01:20:17.200  & $ +$21:33:46.00  & F702W & 2008 &  6115/Zuo   \\
Q$0150$$-$$202$ &01:52:27.300  & $ -$20:01:06.00  & F702W & 5100 &  6557/Steidel  \\
Q$0229$$+$$131$ &02:31:45.894  & $ +$13:22:54.72  & F702W & 5000 &  6557/Steidel \\
Q$0235$$+$$164$ &02:38:38.930  & $ +$16:36:59.28  & F702W & 600  &  5096/Burbidge \\
Q$0349$$-$$146$ &03:51:28.541  & $ -$14:29:08.71  & F702W & 2400 &  5949/Lanzetta \\     
Q$0450$$-$$132$ &04:53:13.556  & $ -$13:05:54.91  & F702W & 2500 &  5984/Steidel \\
Q$0454$$-$$220$ &04:56:08.900  & $ -$21:59:09.00  & F702W & 1200 &  5098/Burbidge \\
Q$0827$$+$$243$ &08:30:52.086  & $ +$24:10:59.82  & F702W & 4600 &  5984/Steidel  \\
Q$0836$$+$$113$ &08:39:33.015  & $ +$11:12:03.82  & F702W & 5000 &  6557/Steidel \\
Q$1019$$+$$309$ &10:22:30.298  & $ +$30:41:05.12  & F702W & 5100 &  6557/Steidel  \\
Q$1038$$+$$064$ &10:41:17.163  & $ +$06:10:16.92  & F702W & 4600 &  5984/Steidel  \\
Q$1127$$-$$145$ &11:30:07.053  & $ -$14:49:27.39  & F814W & 4400 &  9173/Bechtold  \\
Q$1148$$+$$387$ &11:51:29.399  & $ +$38:25:52.57  & F702W & 4800 &  5984/Steidel  \\
Q$1209$$+$$107$ &12:11:40.600  & $ +$10:30:02.00  & F702W & 3600 &  5351/Bergeron \\
Q$1222$$+$$228$ &12:25:27.389  & $ +$22:35:12.72  & F702W & 5000 &  5984/Steidel  \\ 
Q$1241$$+$$176$ &12:44:10.826  & $ +$17:21:04.52  & F702W & 5000 &  6557/Steidel  \\ 
Q$1246$$-$$057$ &12:49:13.800  & $ -$05:59:18.00  & F702W & 4600 &  5984/Steidel  \\
Q$1317$$+$$277$ &13:19:56.316  & $ +$27:28:08.60  & F702W & 4700 &  5984/Steidel  \\
Q$1332$$+$$552$ &13:34:11.700  & $ +$55:01:25.00  & F702W & 2800 &  6557/Steidel  \\
Q$1354$$+$$195$ &13:57:04.437  & $ +$19:19:07.37  & F702W & 2400 &  5949/Lanzetta \\
Q$1424$$-$$118$ &14:27:38.100  & $ -$12:03:50.00  & F702W & 2100 &	 6619/Lanzetta \\
Q$1511$$+$$103$ &15:13:29.319  & $ +$10:11:05.53  & F702W & 5000 &  6557/Steidel  \\
Q$1622$$+$$235$ &16:24:39.090  & $ +$23:45:12.20  & F702W & 24,000 & 5304/Steidel \\
Q$1623$$+$$268$ &16:25:48.793  & $ +$26:46:58.76  & F702W & 4600 &  5984/Steidel  \\
Q$2128$$-$$123$ &21:31:35.262  & $ -$12:07:04.80  & F702W & 1800 &  5143/Mecchetto \\
Q$2206$$-$$199$ &22:08:52.000  & $ -$19:43:59.00  & F702W & 5000 &  6557/Steidel  \\\hline
\end{tabular}}
\end{center}
\end{table*}

Details of the HIRES/Keck and UVES/VLT quasar observations are
presented in Table~\ref{tab:qsospec}. The total integrated exposure
times for each quasar spectrum ranges from 1200 to 53,303 seconds. The
HIRES spectrum of Q$0836+113$ was reduced using the Mauna Kea Echelle
Extraction
(MAKEE\footnote{http://spider.ipac.caltech.edu/staff/tab/makee})
package and the remaining HIRES data were reduced using
IRAF\footnote{IRAF is written and supported by the IRAF programming
group at the National Optical Astronomy Observatories (NOAO) in
Tucson, Arizona. NOAO is operated by the Association of Universities
for Research in Astronomy (AURA), Inc.\ under cooperative agreement
with the National Science Foundation.}.  The UVES data were reduced
using the ESO pipeline and the custom code UVES Post--Pipeline Echelle
Reduction ({\sc uves
popler}\footnote{http://astronomy.swin.edu.au/$\sim$mmurphy/UVES\_popler.html}).
All the quasar spectra are vacuum and heliocentric velocity corrected.
 
The quasar spectra are objectively searched for {\MgII} doublet
candidates using a detection significance level of 5~$\sigma$ for the
$\lambda$2796 line, and 3$\sigma$ for the $\lambda$2803 line.  The
spectra have detection limits of $W_r(2796) \sim 0.02$~{\AA}
(5$\sigma$).  Detections and significance levels follow the formalism
of \citet{schneider93} and \citet{weakI}.  We define a single {\MgII}
system as absorption occurring within $\leq 800$~{\kms}, but this
definition had no affect on our results. A single absorption system
may have several kinematic subsystems \citep[see][]{cv01} --
absorption regions separated by regions of no detected
absorption. Each subsystem is defined in the wavelength regions
between where the per pixel equivalent widths become consistent with
continuum at the $1\sigma$ level.
 
Analysis of the {\MgII} absorption profiles was performed using our
own graphic-based interactive software for local continuum fitting,
objective feature identification and measuring absorption properties.
We compute the equivalent widths, optical depth weighted mean
redshifts ($z_{abs}$), apparent optical depth column densities ($N_a$
[atoms cm$^{-2}$]), and flux decrement weighted velocity centroids,
widths, and asymmetries ($\left< V \right>$, $W_{vs}$, and $A$
[{\kms}]), directly from the normalized pixel counts \citep[see][for
the precise definitions of these quantities]{weakI,archiveI,cv01}.
Quantities for the full absorption systems are measured between the
pixel of the most blueward extreme and redward extreme subsystems
while omitting pixels consistent with continuum (those outside/between
the kinematic subsystems).
 
In many of the {\MgII} systems, corresponding transitions from other
elements (such as {\MgI} $\lambda$2853, {\FeII} $\lambda$2344,
$\lambda$2374, $\lambda$2383, $\lambda$2587, $\lambda$2600, {\CaII}
$\lambda\lambda$3935, 3970, and {\MnII} $\lambda$2577, $\lambda$2594,
$\lambda$2606) are detected, which further validates the
identification of a {\MgII} system, but are not required to validate a
system. In Figure~\ref{fig:satprofile} we show an example {\MgII}
system that has corresponding transitions from other elements.
 
In addition to the aforementioned quantities measured directly from
the flux decrements, we modeled the absorption profiles using Voigt
profile least--square fitting.  Voigt profile decomposition provides a
means to model each complex absorption profile as multiple individual
isothermal ``clouds''.  Each cloud is parametrized by a velocity
center, a column density, and Doppler parameter.  The overall
decomposition provides a model of the number of clouds ($N_{cl}$) and
their individual Voigt profile parameters. 

\setcounter{figure}{0}
\begin{figure*}
\includegraphics[angle=0,scale=0.43]{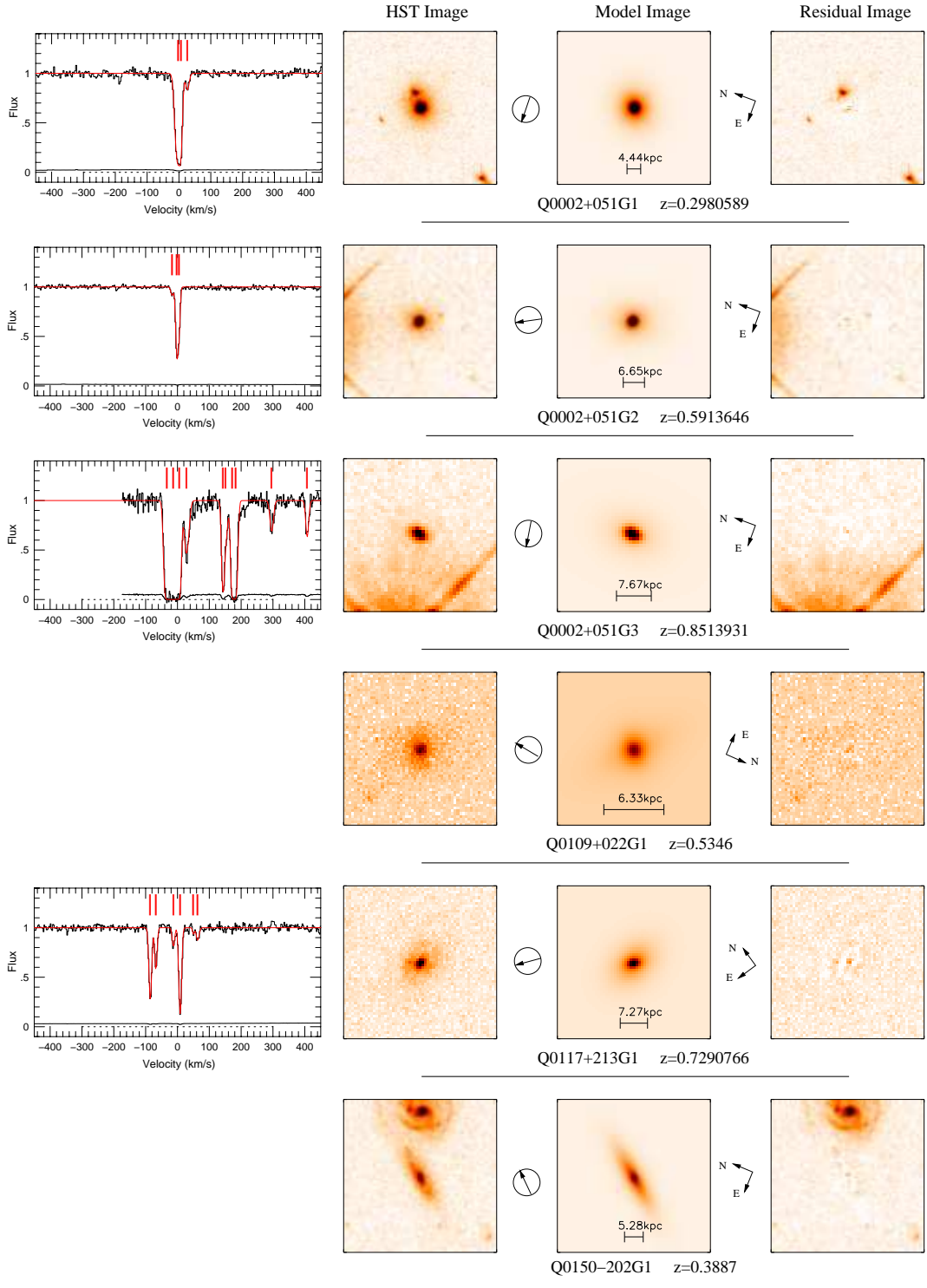}
\caption{--- The full version of this Figure is in the electronic version of the paper, not in the printed version. 
In the printed version we present a subset of our full sample. --- (far-left) The HIRES/Keck or UVES/VLT quasar spectra of
the {{\rm Mg}\kern 0.1em{\sc ii}~$\lambda 2796$} absorption feature
are shown alongside the associated absorbing galaxy on the right. The
{\MgII} $\lambda 2796$ optical depth weight mean absorption redshift
is the zeropoint of the velocity scale. The tick marks indicate the
number of Voigt profile components and the red curve indicates the fit
to the data.  We do not have HIRES or UVES data available for six
galaxies. (left) WFPC-2/{\it HST} images of galaxies selected by
{\MgII} absorption. The images are 10 times larger than the
$1.5\sigma$ isophotal area.  --- (center) The GIM2D models of the
galaxies, which provide quantified morphological parameters. A scale
of one arcsecond is indicated on each image along with the physical
scale computed at the {\MgII} absorption redshift.  --- (right) The
residual images from the models, showing quality of the fit and the
underlying structure and morphological perturbations of the
galaxies. The encircled arrow provides the direction to the quasar
(galaxy--quasar orientation). The cardinal directions are also shown
and the quasar name and redshift of {\MgII} absorption is provided
under each set of galaxy WFPC-2, model and residual image.}
\label{fig:mos}
\end{figure*}

To best constrain the model, the profile fits simultaneously
incorporate all transitions associated with the {\MgII} system.  This
is particularly useful in cases where the {\MgII} absorption is
saturated and there is a loss of component structure.  Thus, we use
the unsaturated {\FeII} and/or {\CaII} and/or {\MgI} transitions to
constrain the number of clouds and velocity centers.  All transitions
have the same number of clouds and the velocities of the clouds are
tied for all transitions.  For this work, we assume thermally
broadened clouds (no turbulent component), so the $b$ parameters of
each cloud are tied for all transitions (yielding the broadening
temperature to be the same for all ions).  The column densities of
transitions belonging to a given ion are tied, but the column
densities of different ions are freely fit.
 
In Figure~\ref{fig:satprofile}, we show the least--squares Voigt
profile model (red) of a saturated {\MgII} doublet with associated
transitions from other ions.  The number of individual Voigt profile
components ($N_{cl}$), or ``clouds'', are indicated with vertical tick
marks.
 
The fitting philosophy is to obtain the minimum number of clouds that
provide a statistically reasonable model (i.e., $\chi_\nu \simeq 1$).
We use the code MINFIT \citep{cwcthesis}, which takes a user input
model and performs a least--squares fit while minimizing the number of
clouds.  With each iteration, the least statistically significant
cloud (based upon fractional error criterion) is removed from the
model and a new model (less the one cloud) is fit.  An F--test is
performed to determine whether the two models are statistically
consistent at the 97\% confidence level.  If they are consistent, the
cloud is removed from the model.  Each cloud not meeting the
fractional error criterion is tested in this way until the all clouds
meet the criterion and the number of clouds is minimized via the
F--test iterations.  We applied the fractional error criterion
$\sqrt{(dN/N)^2 + (db/b)^2 + (dz/z)^2} < 1.5 $, where $dN/N$, $db/b$,
and $dz/z$ are the column density, $b$ parameter, and redshift
fractional errors of the cloud.  The resulting models are not
sensitive to the input model as long as the input model contains a
sufficient number of clouds.  For further description of the Voigt
profile decomposition of {\MgII} systems see \citet{cwcthesis},
\citet{cv01}, and \citet{cvc03}.
 
We compute a total system Voigt profile column density,
$N_{\hbox{\tiny VP}}$ [atoms cm$^{-2}$], and its uncertainty by
summing the individual cloud column densities over the total number of
clouds.  This number is comparable to the apparent optical depth
column density in unsaturated systems, but provides a slightly more
robust total column density in saturated systems where the apparent
column density provides only a lower limit.

\begin{figure}
\includegraphics[angle=0,scale=1.20]{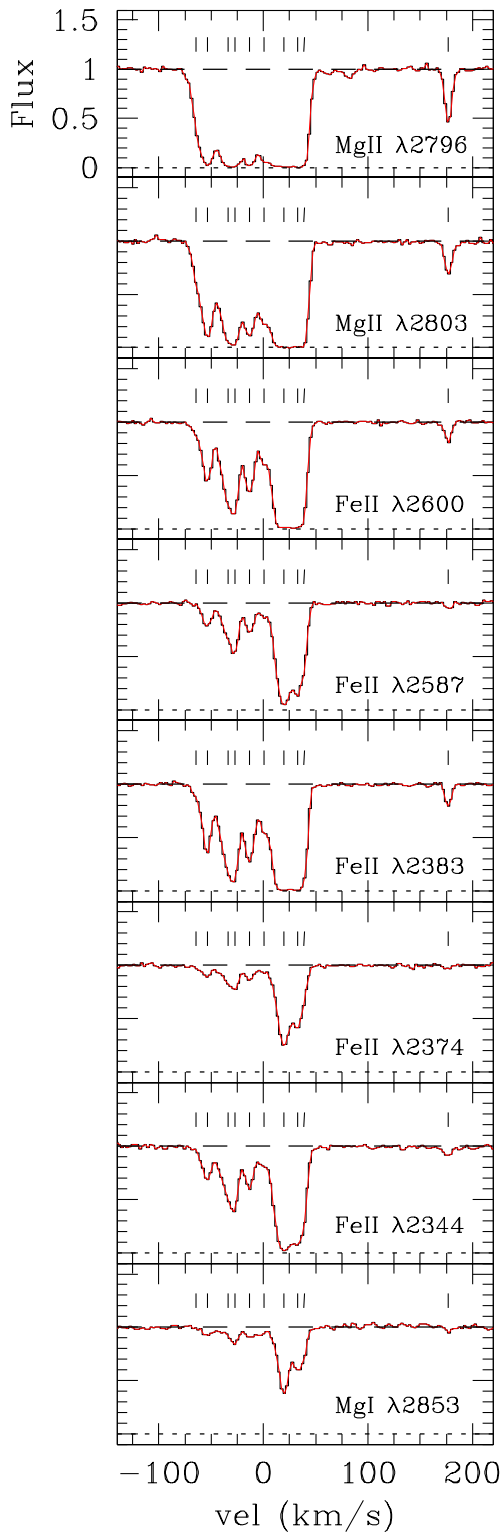}
\caption{The saturated {\MgII} doublet, {\FeII} and {\MgI} are shown
for $z_{abs}=1.017038$ absorption system seen in the quasar
Q2206--199.  The Voigt profile fits (red) and the number of individual
Voigt profile components ($N_{cl}$) are indicated with vertical tick
marks.  The {\FeII} and {\MgI} profiles are used to constrain the
kinematics of the saturated {\MgII} profile and to derive {\MgII}
column densities.}
\label{fig:satprofile}
\end{figure}

\subsection{{\it \bf HST} Imaging of Quasar Fields}

Details of the WFPC-2/{\it HST\/} quasar field observations are
presented in Table~\ref{tab:qsoHST}. The WFPC--2/{\it HST\/} F702W and
F814W images were reduced using the WFPC--2 Associations Science
Products Pipeline (WASPP\footnote{Developed by the Canadian Astronomy
Data Centre (CADC) and the Space Telescope--European Coordinating
Facility (ST--ECF): {\it
http://archive.stsci.edu/hst/wfpc2/pipeline.html}}).  WASPP data
quality verifications include photometric and astrometric accuracy and
correctly set zero--points. The F702W filter provides a bandpass
similar to a rest-frame Johnson $B$--band filter for galaxies at
$z\sim 0.6$. The F814W filter provides a bandpass similar to the
rest-frame $B$--band for galaxies at $z\sim 0.85$.  Galaxy photometry
was performed using the Source Extractor (Sextractor) package
\citep{bertin96} with a detection criterion of 1.5~$\sigma$ above
background with a minimum object area 75 pixels.  The $m_{F702W}$ and
$m_{F814W}$ magnitudes were measured using the WFPC--2/{\it HST\/}
zero points \citep{whitmore95}, based upon the Vega system.

Galaxy absolute $B$-band magnitudes, $M_B$, were determined from the
$k$--corrected observed $m_{F702W}$ or $m_{F814W}$.  To compute the
$k$--corrections, we adopted spectral energy distribution (SED)
templates of \citet{kinney96}.  The adopted SED for each galaxy was
based upon its rest-frame $B-K$ color obtained from \citet{sdp94}. For
galaxies with no color information, we adopted an Sb SED since this is
consistent the with average color of a {\MgII} absorbing galaxy
\citep{sdp94,zibetti07}.  $B$--band luminosities were computed using
the DEEP2 optimal $M^{\ast}_B$ of \citet[][Table~2]{faber07} in the
redshift bin appropriate for each galaxy: $M^{\ast}_B$ ranges from
$-21.07$ ($\left< z \right> =0.3$) to $-21.54$ ($\left< z\right>
=1.1$).

\subsection{GIM2D Galaxy Models}

For each galaxy, the morphological parameters were quantified by
fitting a two-component (bulge+disk) co-spatial parametric model to
its two-dimensional surface brightness distribution using GIM2D
(Galaxy IMage 2D) \citep{simard02}.  GIM2D uses the Metropolis
algorithm \citep{metropolis53,saha94}, which does not easily fall prey
to local minima, to explore the complicated topology of this twelve
dimensional parameter space.  Once a convergence point is satisfied,
the algorithm Monte-Carlo samples the region and keeps acceptable
parameter sets, building up a solution probability distribution, which
it uses to compute the median of the probability distribution of each
free parameter and its 1$\sigma$ uncertainties. This technique results
in an optimal model parameter set and asymmetric errors.

For our 40 galaxies, we fit the surface brightness of the disk
component with an exponential profile and we fit the bulge component
with a S{\'e}rsic profile \citep{sersic68} where the S{\'e}rsic index
may vary between $0.2\leq n\leq 4.0$. These model fits differ from
\citet{kacprzak07} where we focus was on galaxy asymmetries and how
absorption-selected galaxies differ from an ideal galaxy with a de
Vaucouleurs bulge and an exponential disk profile. The models have a
maximum of twelve free parameters:

\newcounter{qcounter}
\begin{list}{~~~~~~~~~~~~~~~~~~~~(\arabic{qcounter}.)~}{\usecounter{qcounter}}
\item ~~Galaxy total flux
\item ~~Bulge-to-total fraction ($B/T$)
\item ~~Bulge semi-major axis effective radius ($r_b$)
\item ~~Bulge ellipticity ($e_b\equiv 1-b/a$)
\item ~~Bulge position angle ($\theta_b$) 
\item ~~Bulge S{\'e}rsic index ($n$) 
\item ~~Disk scale length ($r_d$)
\item ~~Disk inclination ($i$)
\item ~~Disk position angle ($\theta_d$)
\item Model center sub-pixel offsets ($dx$) 
\item Model center sub-pixel offsets ($dy$) 
\item Background residual level.
\end{list}

A PSF-deconvolved half-light semi-major axis radius $r_h$ is also
computed for each galaxy by integrating the sum of bulge and disk
surface brightness profiles out to infinity using the best fitting
structural parameters.  The half-light radius may be unreliable for
cases where there are large differences between the bulge and disk
position angles.  Additional information regarding the structural
parameters derived by GIM2D can be obtained from \citet{simard02}.
GIM2D outputs the various scale length parameters in units of pixels
and we use the plate scale of the appropriate WFPC-2 chip to convert
to an angular quantity. We then use the plate scale and the angular
diameter distance from the adopted cosmology to convert them to linear
quantities.

GIM2D uses the isophotal area to extract ``portrait size'' galaxy
images from original images with an area 10 times larger than the
$1.5\sigma$ galaxy isophotal area, which is chosen such that an
accurate background can be computed by GIM2D.  During the GIM2D
modeling process, the models are convolved with the WFPC-2 point
spread function, which was modeled at the appropriate locations on the
WFPC-2 chip using Tiny Tim \citep{krist04}.

The GIM2D outputs were manually inspected to see if models were
realistic representations of the observed galaxies.  In addition, we
used previously measured galaxy rotation curves to further validate
several galaxy models \citep{kacprzak10a,steidel02}. In 4/40 cases, we
found that the model settled on an unreasonable solution (galaxies
Q0454G1, Q1148G1, Q1354G1, and Q2206G2).  We reran GIM2D for these
four systems using different random seeds, which resulted in more a
realistic and stable model.  A GIM2D fail-rate of 10\% may be less
important for large samples of galaxies, however with our smaller
sample it is crucial to ensure that we produce representative models.

We note that galaxy asymmetries, and structures such as tidal tails
and bars, may affect the galaxy isophotal shape, thereby providing an
under/over estimate of the true galaxy shape and orientation.  It is
difficult to model these disturbed systems and they are not
straightforward to interpret. Thus, in cases such as the interacting
pair Q0450G1 and Q0450G2, the model inclinations are likely to be
biased with respect to the real physical inclination of the disk.
Thus, the model results presented here should be interpreted with
some caution. However, despite the simplicity of the two-component
model (i.e., real galaxies may have more than two smooth components,
such as spiral arms, bars, {\HII} regions, etc.), careful analysis of
the output models and parameters provides useful information regarding
the complexity of galaxies \citep[see][]{kacprzak07}.

The impact parameters, $D$, and their uncertainties are computed using
a combination of galaxy isophotal centroids determined by SExtractor
and GIM2D model offsets.  The dominant uncertainty in $D$ is derived
from the pixel offset of the galaxy isophotal center obtained using
SExtractor and the isophotal center of the galaxy model determined by
GIM2D.  This offset is typically about $0.25$ pixels. There is also a
$\sim 0.05$ pixel uncertainty in the position of the quasar based upon
centroiding errors of unresolved sources in the images.

\begin{figure}
\includegraphics[angle=0,scale=0.45]{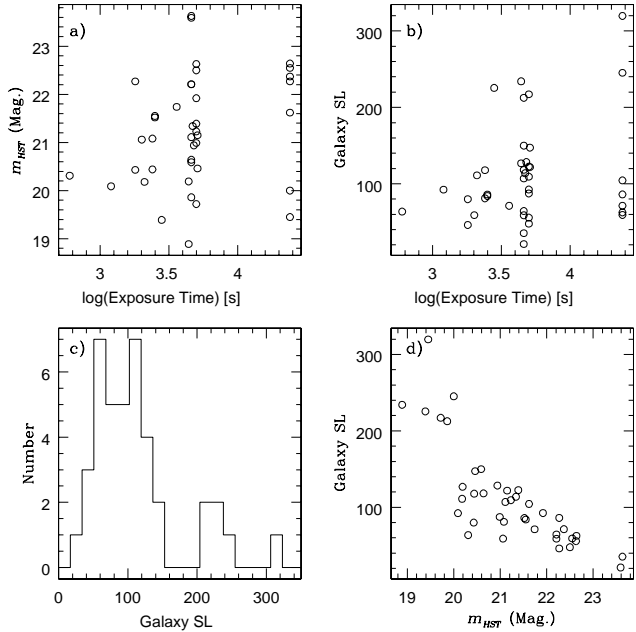}
\caption{--- (a) The apparent magnitude as a function of exposure time
for the 40 {\MgII} absorption-selected galaxies in our sample. --- (b)
Galaxy significance level ($SL$) as a function of exposure time. ---
(c) Galaxy $SL$ distribution.  --- (d) Galaxy $SL$ as a function of
galaxy apparent magnitude. Note that the galaxy $SL$ is only dependent
$m_{HST}$ and not the WFPC-2 exposure time.}
\label{fig:chp2SL}
\end{figure}

\subsubsection{Galaxy Significance Levels and GIM2D Models}

The WFPC-2 image exposure times for our sample range from 600--24,000
seconds, with a typical time of $\sim$4700 seconds. It may be of
concern that the longest and shortest exposures could yield different
measured morphological parameters for similar type galaxies.  However,
the problem is not so simple since there can be several galaxies of
interest in each quasar field image having different apparent
magnitudes.  \citet{simard02} studied sets of GIM2D simulations to
characterize the systematic biases and random errors in the galaxy
structural measurements. They analyzed simulated WFPC-2 F814W images
containing $\sim$5500 simulated galaxies.  Sky photon noise and
detector read-out noise were included along with background noise
brightness fluctuations contributed from very faint galaxies below the
detection threshold. \citet{simard02} determined that the structural
parameters can be recovered from the simulated images, equivalent to
typical WFPC-2 exposure times of $\sim$2800--4400~s, and that the
uncertainties only become significant when $m_{F814W} > 23.5$.  Only
two of our galaxies have apparent magnitudes above the Simard limit
with the remaining below $m_{HST}=22.6$. Both of those galaxies have
$m_{HST}=23.6$ and are in a WFPC-2 image that has a typical exposure
time of 5000~seconds.

Here we examine the significance level ($SL$) for each galaxy in our
sample, which is defined by the ratio of the galaxy measured flux and
the flux uncertainty (see Table~\ref{tab:galmorph}). The $SL$ is then
equivalent to an average signal-to-noise ratio per pixel, with higher
weighting towards the bright pixels. In Figure~\ref{fig:chp2SL}$a$, we
show the distribution of galaxy apparent magnitudes as a function of
exposure time. All exposure times cover a wide range of apparent
magnitudes. We see no trends with exposure time.

In Figure~\ref{fig:chp2SL}$b$, we show the distribution of $SL$ as a
function of exposure time. There is a wide range of $SL$ for each
exposure time.  In Figure~\ref{fig:chp2SL}$c$, we show the binned
distribution of $SL$. The majority of galaxies lie within the main
peak around $SL \sim 80$ with others, not necessarily from long
exposures, at higher $SL$. In Figure~\ref{fig:chp2SL}$d$, we show the
$SL$ as a function of galaxy apparent magnitude. Note the strong
anti-correlation. The anti-correlation exhibits some dispersion which
can be attributed to background Poisson noise and also some variance
in the galaxy surface brightness for a given magnitude.

The strong anti-correlation between $SL$ and $m_{HST}$ suggests that
the galaxy $SL$ primarily depends on its apparent magnitude and not on
the exposure time.  Therefore, degrading individual quasar field
images to a lower signal-to-noise ratio is not necessary for the
galaxy magnitude range of interest ($m_{HST} < 23.5$). We discuss
additional tests regarding $S/N$ and quasar PSF near absorbing
galaxies in \citet{kacprzak07}.

\begin{table*}
\begin{center}
  \caption{The observed galaxy properties. The columns are (1) the
  quasar name and galaxy ID number, (2) the optical depth weighted
  mean {\MgII} $\lambda2796$ absorption redshift, (3) the {\MgII}
  $\lambda2796$ rest-frame equivalent width, (4) the source of the
  adopted equivalent width measurements, (5) the adopted galaxy
  redshift, (6) the source(s) of galaxy spectroscopic redshift, (7)
  the projected galaxy--quasar separation (impact parameter), (8) the
  WFPC-2/{\it HST} F702W or F814W galaxy apparent magnitude, (9)
  galaxy significance level in the WFPC--2 images, defined to be the
  ratio of the measured galaxy flux and the flux uncertainty, (10) the
  $B-K$ colors, (11) the rest-frame $B$-band galaxy absolute
  magnitude, and (12) the galaxy $B$-band luminosity. }
  \vspace{-0.5em}
\label{tab:galmorph} 
{\footnotesize\begin{tabular}{llccllrcrrcc}\hline
QSO field&\phantom{00}$z_{abs}$& $W_r(2796)$  & Ref$^{ a}$&\phantom{0}$z_{gal}$&Ref$^{ b}$& $D$\phantom{0000} & $m_{HST}$ &  $SL$\phantom{0}& $B$$-$$K$&$M_B$& $L_B$ \\
\& Galaxy ID       &            &(\AA)           &              &     &  & (kpc)\phantom{00}  &                &             &     &       &  ($L^{\star}$) \\\hline
 Q$0002$$+$$051$G1 & $0.298059$ & $0.244\pm0.003$ & 1 & $0.298  $ & 1    & $59.7\pm0.3$ & $19.86\pm0.01$ & 212.6 & $4.02$ & $-19.64$ & $0.27$ \\ 
 Q$0002$$+$$051$G2 & $0.591365$ & $0.102\pm0.002$ & 1 & $0.592  $ & 1    & $36.3\pm0.3$ & $21.11\pm0.01$ & 106.9 & $4.12$ & $-20.78$ & $0.71$ \\ 
 Q$0002$$+$$051$G3 & $0.851393$ & $1.089\pm0.008$ & 1 & $0.85180$ & 2    & $26.0\pm0.4$ & $22.21\pm0.02$ & 64.4  & $2.86$ & $-21.27$ & $0.92$ \\ 
 Q$0109$$+$$200$G1 & $0.5346  $ & $2.26$\phantom{0000.008} & 2 & $0.534  $ & 3,4  & $45.1\pm0.1$ & $22.27\pm0.02$ & 46.1 & $3.90$ & $-19.20$ & $0.17$ \\ 
 Q$0117$$+$$213$G1 & $0.729077$ & $0.244\pm0.005$ & 1 & $0.729  $ & 1    & $55.8\pm0.3$ & $21.06\pm0.02$& 59.0  & $4.02$ & $-21.81$ & $1.32$ \\ 
 Q$0150$$-$$202$G1 & $0.3887  $ & $0.58 \pm0.05 $ & 2 & $0.383  $ & 3,4  & $60.6\pm0.7$ & $21.15\pm0.01$& 121.8 &$\cdots$& $-19.26$ & $0.19$ \\ 
 Q$0229$$+$$131$G1 & $0.417338$ & $0.816\pm0.020$ & 1 & $0.4167 $ & 2,3,4& $37.1\pm0.5$ & $19.72\pm0.01$& 217.1 & $3.92$ & $-20.83$ & $0.74$ \\ 
 Q$0235$$+$$164$G1 & $0.524   $ & $2.34\pm0.05$   & 3 & $0.524  $ & 4    & $12.2\pm0.2$ & $20.31\pm0.02$& 63.6  &$\cdots$& $-21.10$ & $0.95$ \\ 
 Q$0349$$-$$146$G1 & $0.357168$ & $0.175\pm0.007$ & 1 & $0.3567 $ & 5    & $71.9\pm0.3$ & $20.44\pm0.01$& 117.8 &$\cdots$& $-19.70$ & $0.28$ \\ 
 Q$0450$$-$$132$G1 & $0.493936$ & $0.674\pm0.024$ & 1 & $0.4941 $ & 1,2  & $50.1\pm0.1$ & $21.55\pm0.01$& 84.0  & $3.41$ & $-19.65$ & $0.25$ \\ 
 Q$0450$$-$$132$G2 & $0.493936$ & $0.674\pm0.024$ & 1 & $0.4931 $ & 1,2  & $62.8\pm0.1$ & $21.52\pm0.01$& 86.0  & $3.41$ & $-19.69$ & $0.26$ \\ 
 Q$0454$$-$$220$G1 & $0.483337$ & $0.426\pm0.007$ & 1 & $0.48382$ & 2,5  & $107.8\pm0.8$ & $20.09\pm0.01$& 92.3 &$\cdots$& $-21.03$ & $0.90$ \\ 
 Q$0827$$+$$243$G1 & $0.524966$ & $2.419\pm0.012$ & 1 & $0.5247 $ & 6    & $37.5\pm0.3$ & $20.64\pm0.01$& 118.2 & $4.01$ & $-20.75$ & $0.69$ \\ 
 Q$0836$$+$$113$G1 & $0.786725$ & $2.133\pm0.019$ & 1 & $0.78682$ & 2,7  & $26.9\pm0.5$ & $22.63\pm0.02$& 55.8  & $3.00$ & $-20.48$ & $0.39$ \\ 
 Q$1019$$+$$309$G1 & $0.3461  $ & $0.70 \pm0.07$  & 4 & $0.346$   & 1    & $46.3\pm0.1$ & $20.46\pm0.01$& 147.3 & $3.10$ & $-19.59$ & $0.26$ \\ 
 Q$1038$$+$$064$G1 & $0.441453$ & $0.673\pm0.011$ & 1 & $0.4432 $ & 8  & $56.2\pm0.3$ & $20.59\pm0.01$ & 149.9  & $4.43$ & $-20.15$ & $0.40$ \\ 
 Q$1127$$-$$145$G1 & $0.328279$ & $0.028\pm0.003$ & 1 & $0.32839$ & 9  & $76.9\pm0.4$ & $20.19\pm0.01$ & 126.8  &$\cdots$& $-19.16$ & $0.17$ \\ 
 Q$1127$$-$$145$G2 & $0.328279$ & $0.028\pm0.003$ & 1 & $0.32847$ & 2  & $91.4\pm0.2$ & $18.89\pm0.00$ & 234.1  &$\cdots$& $-20.46$ & $0.57$ \\ 
 Q$1148$$+$$387$G1 & $0.553363$ & $0.640\pm0.013$ & 1 & $0.5536 $ & 8  & $20.6\pm0.3$ & $20.94\pm0.01$ & 128.4  & $3.59$ & $-20.68$ & $0.65$ \\ 
 Q$1209$$+$$107$G1 & $0.392924$ & $1.187\pm0.005$ & 1 & $0.392  $ & 3  & $37.9\pm0.1$ & $21.74\pm0.02$ & 71.2   & $1.89$ & $-18.60$ & $0.10$ \\ 
 Q$1222$$+$$228$G1 & $0.550198$ & $0.094\pm0.009$ & 1 & $0.5502 $ & 8  & $38.0\pm0.6$ & $22.50\pm0.02$ & 47.7   & $3.80$ & $-19.08$ & $0.15$ \\ 
 Q$1241$$+$$176$G1 & $0.550482$ & $0.465\pm0.011$ & 1 & $0.550  $ & 1  & $21.2\pm0.3$ & $21.39\pm0.01$ & 122.5  & $3.42$ & $-20.21$ & $0.42$ \\ 
 Q$1246$$-$$057$G1 & $0.639909$ & $0.450\pm0.004$ & 1 & $0.637$ & 1  & $29.3\pm0.9$ & $22.21\pm0.02$ &  58.8    & $3.78$ & $-20.02$ & $0.25$ \\ 
 Q$1317$$+$$277$G1 & $0.660049$ & $0.320\pm0.006$ & 1 & $0.6610  $ & 8  & $103.9\pm0.5$ & $21.34\pm0.01$& 113.8 &$\cdots$& $-20.97$ & $0.61$ \\ 
 Q$1332$$+$$552$G1 & $0.374   $ & $2.90$\phantom{0000.008}  & 2 & $0.373  $ & 10  & $28.0\pm0.3$ & $19.39\pm0.00$& 225.5 & $3.95$ & $-20.79$ & $0.77$ \\ 
 Q$1354$$+$$195$G1 & $0.456598$ & $0.773\pm0.015$ & 1 & $0.4592 $ & 11 & $45.3\pm0.5$ & $21.08\pm0.01$  & 81.0 & $3.04$ & $-19.85$ & $0.30$ \\ 
 Q$1424$$-$$118$G1 & $0.341716$ & $0.100\pm0.015$ & 1 & $0.3404 $ & 12 & $86.8\pm0.4$ & $20.18\pm0.01$  & 111.1&$\cdots$& $-19.83$ & $0.32$ \\ 
 Q$1511$$+$$103$G1 & $0.4369  $ & $0.454\pm0.046$ & 5 & $0.437  $ & 3,4& $38.3\pm0.4$ & $21.23\pm0.01$  & 109.3 & $3.02$ & $-19.56$ & $0.23$ \\ 
 Q$1622$$+$$235$G1 & $0.317597$ & $0.491\pm0.010$ & 1 & $0.3181$ & 13  & $54.8\pm0.2$ & $20.00\pm0.00$  & 245.2 & $4.09$ & $-19.68$ & $0.28$ \\ 
 Q$1622$$+$$235$G2 & $0.368112$ & $0.247\pm0.005$ & 1 & $0.3675 $ &13  & $114.5\pm0.1$ & $19.45\pm0.00$ & 319.7 &$\cdots$& $-20.79$ & $0.77$ \\ 
 Q$1622$$+$$235$G3 & $0.471930$ & $0.769\pm0.006$ & 1 & $0.4720$ & 13  & $34.3\pm0.2$ & $22.27\pm0.01$  & 86.1  & $2.71$ & $-18.76$ & $0.11$ \\ 
 Q$1622$$+$$235$G4 & $0.656106$ & $1.446\pm0.006$ & 1 & $0.6560$ & 13  & $100.0\pm0.7$ & $22.55\pm0.02$ & 59.2  & $2.49$& $-19.71$ & $0.19$ \\ 
 Q$1622$$+$$235$G5 & $0.702902$ & $0.032\pm0.003$ & 1 & $0.7016$ & 13  & $113.2\pm0.5$ & $21.62\pm0.01$& 104.4  &$\cdots$& $-20.97$ & $0.61$ \\ 
 Q$1622$$+$$235$G6 & $0.797078$ & $0.468\pm0.008$ & 1 & $0.7975$ & 13  & $71.8\pm0.7$ & $22.37\pm0.02$  &  71.3 & $3.04$ & $-20.80$ & $0.52$ \\ 
 Q$1622$$+$$235$G7 & $0.891276$ & $1.548\pm0.004$ & 1 & $0.8909$ & 13  & $23.4\pm0.2$ & $22.64\pm0.02$ &  62.6  & $3.10$ & $-21.04$ & $0.74$ \\ 
 Q$1623$$+$$268$G1 & $0.887679$ & $0.903\pm0.004$ & 1 & $0.888$   & 1  & $48.2\pm0.5$ & $23.63\pm0.03$ &  35.4  & $3.97$ & $-20.29$ & $0.37$ \\ 
 Q$1623$$+$$268$G2 & $0.887679$ & $0.903\pm0.004$ & 1 & $0.888$   & 1  & $72.0\pm0.2$ & $23.59\pm0.05$ &  21.2  &$\cdots$& $-20.07$ & $0.30$ \\ 
 Q$2128$$-$$123$G1 & $0.429735$ & $0.395\pm0.010$ & 1 & $0.430  $ & 3,4  & $48.4\pm0.2$ & $20.43\pm0.01$&  79.9 & $3.26$ & $-20.32$ & $0.46$ \\ 
 Q$2206$$-$$199$G1 & $0.948361$ & $0.249\pm0.002$ & 1 & $0.948  $ & 3   & $87.6\pm0.7$ & $21.92\pm0.01$ &  92.5 &$\cdots$& $-22.02$ & $1.83$ \\
 Q$2206$$-$$199$G2 & $1.017038$ & $1.047\pm0.003$ & 1 & $1.01655$ & 3,2  & $105.2\pm0.6$& $20.99\pm0.01$&  87.4 &$\cdots$& $-23.79$ & $7.96$ \\\hline 
\end{tabular}}
\end{center}
$^a${\MgII} Absorption Measurements:(1)~This paper (2)~\citet{gb97}, (3)~\citet{lanzetta92}, (4)~\citet{steidel92}, and 
(5)~\citet{foltz86}.\\
$^b$Galaxy Identification:
(1)~\citet{sdp94}, (2)~\citet{kacprzak10a}, (3)~\citet{bb91},
(4)~\citet{gb97}, (5)~\citet{chen98}, (6)~\citet{kanekar01},
(7)~\citet{lowenthal90}, (8)~\citet{steidel02}, (9)~\citet{kacprzak10b},
(10)~\citet{miller87}, (11)~\citet{ellingson91}, (12)~\citet{chen01}, and (13)~\citet{s97}.

\end{table*}
\begin{table*}
\begin{center}
  \caption{Summary of the {\MgII} $\lambda 2796$ absorption
properties.  The table columns are (1) the quasar name and galaxy ID
number, (2) the optical depth weighted mean {\MgII} $\lambda2796$
absorption redshift, (3) the rest-frame equivalent width $W_r(2796$),
(4) the doublet ratio ($DR$), (5) the number of clouds ($N_{cl}$), (6)
the Voigt profiles fitted system column density ($N_{vp}$), (7) the
{\MgII} optical depth, (8) The AOD derived column density ($N_a$), (9)
the mean velocity ($\left< V \right>$), (10) the velocity spread,
($W_{vs}$), and (9) the velocity asymmetry ($A$). }
  \vspace{-0.5em}
\label{tab:abs}
{\footnotesize\begin{tabular}{llccrlrcrrrr}\hline
QSO field & \phantom{000}$z_{abs}$& $W_r(2796)$           &    $DR$      &$N_{cl}$&\phantom{0}log($N_{vp}$)& $\tau$$^a$\phantom{000}                 &      log$(N_a)$$^a$              &     $\left< V \right>$\phantom{00} &$W_{vs}$\phantom{00}       &    $A$ \phantom{000}       \\
\& Galaxy ID&          &             (\AA)     &                    &        & \phantom{0.}(cm$^{-2}$)       &                    &       (cm$^{-2}$)    &      (\kms)                &       (\kms)         &        (\kms) \\\hline
Q$0002$$+$$051$G1 & $0.298058$ & $0.244$$\pm$$0.003$  &$1.34$$\pm$$0.03$ &    3   & $13.14$$\pm$$0.02$ & 54.1$^{+8.6}_{-9.9}$   & 13.08$^{+0.07}_{-0.08}$  & $0.12$$\pm$$0.33$  & $11.6$$\pm$$0.4$  & $0.33$$\pm$$0.06$ \\ [1.0ex]
Q$0002$$+$$051$G2 & $0.591363$ & $0.102$$\pm$$0.002$  &$1.54$$\pm$$0.04$ &    3   & $12.61$$\pm$$0.04$ & 16.0$^{+3.0}_{-3.2}$   & 12.55$^{+ 0.08}_{-0.09}$ & $-0.39$$\pm$$0.29$ & $7.0$$\pm$$0.5$   & $-0.41$$\pm$$0.25$ \\[1.0ex] 
Q$0002$$+$$051$G3 & $0.851349$ & $1.089$$\pm$$0.008$  &$1.16$$\pm$$0.01$ &   10   & $14.43$$\pm$$0.04$ & 406.8$^{+22.1}_{-22.8}$& 13.95$^{+ 0.02}_{-0.02}$ & $78.36$$\pm$$1.20$ & $112.5$$\pm$$1.5$ & $0.95$$\pm$$0.02$ \\ [1.0ex]
Q$0109$$+$$200$G1 & $0.5346  $ & $2.26 $\phantom{000.000} & $ \cdots$    &$\cdots$& $ \cdots$            & $ \cdots$            & $ \cdots$                & $ \cdots$          & $ \cdots$         & $ \cdots$       \\ [1.0ex]
Q$0117$$+$$213$G1 & $0.729104$ & $0.244$$\pm$$0.005$  &$1.84$$\pm$$0.09$ &    6   & $13.05$$\pm$$0.02$ & 38.5$^{+4.3}_{-5.1}$   & 12.93$^{+ 0.05}_{-0.06}$ & $-29.54$$\pm$$1.28$& $48.5$$\pm$$1.0$  & $0.86$$\pm$$0.21$ \\ [1.0ex]
Q$0150$$-$$202$G1 & $0.388700$ & $0.580$$\pm$$0.050$  & $ \cdots$        &$\cdots$& $ \cdots$            & $ \cdots$            & $ \cdots$                & $ \cdots$          & $ \cdots$         & $ \cdots$       \\ [1.0ex]
Q$0229$$+$$131$G1 & $0.417316$ & $0.816$$\pm$$0.020$  &$1.16$$\pm$$0.04$ &    6   & $13.84$$\pm$$0.03$ & 186.4$^{+9.5}_{-19.8}$ & 13.61$^{+ 0.02}_{-0.05}$ & $8.56$$\pm$$2.91$  & $64.7$$\pm$$5.6$  & $1.08$$\pm$$0.08$ \\ [1.0ex]
Q$0235$$+$$164$G1 & $0.524   $ & $2.34 $$\pm$$0.05$   &$ \cdots$         &$\cdots$& $ \cdots$            & $ \cdots$            & $ \cdots$                & $ \cdots$          & $ \cdots$        & $ \cdots$          \\ [1.0ex]
Q$0349$$-$$146$G1 & $0.357161$ & $0.175$$\pm$$0.007$  &$1.17$$\pm$$0.07$ &    3   & $13.87$$\pm$$0.17$ & 38.0$^{+6.2}_{-8.9}$   & 12.92$^{+ 0.07}_{-0.10}$ & $-24.47$$\pm$$2.20$& $43.5$$\pm$$2.1$  & $-2.48$$\pm$$0.13$ \\ [1.0ex]
Q$0450$$-$$132$G1 & $0.493929$ & $0.674$$\pm$$0.024$  &$1.19$$\pm$$0.06$ &    6   & $14.74$$\pm$$2.77$ & $>$114.6               & $>$13.70                 & $-1.86$$\pm$$1.19$ & $24.3$$\pm$$1.2$  & $-0.20$$\pm$$0.07$ \\[1.0ex] 
Q$0450$$-$$132$G2 & $0.493929$ & $0.674$$\pm$$0.024$  &$1.19$$\pm$$0.06$ &    6   & $14.74$$\pm$$2.77$ & $>$114.6               & $>$13.70                 & $-1.86$$\pm$$1.19$ & $24.3$$\pm$$1.2$  & $-0.20$$\pm$$0.07$ \\ [1.0ex]
Q$0454$$-$$220$G1 & $0.483319$ & $0.426$$\pm$$0.007$  &$1.33$$\pm$$0.04$ &    6   & $13.69$$\pm$$0.10$ & $>$70.6                & $>$13.49                 & $-4.19$$\pm$$0.59$ & $19.6$$\pm$$0.7$  & $-0.34$$\pm$$0.04$ \\ [1.0ex]
Q$0827$$+$$243$G1 & $0.524988$ & $2.419$$\pm$$0.012$  &$1.04$$\pm$$0.01$ &   13   & $15.25$$\pm$$0.05$ & $>$640.5               & $>$14.45                 & $4.20$$\pm$$0.50$  & $78.7$$\pm$$0.4$  & $0.09 $$\pm$$0.02$ \\ [1.0ex]
Q$0836$$+$$113$G1 & $0.786731$ & $2.113$$\pm$$0.019$  &$1.05$$\pm$$0.01$ &   20   & $15.54$$\pm$$2.29$ & $>$527.2               & $>$14.37                 & $0.95$$\pm$$0.70$  & $67.5$$\pm$$0.5$  & $0.13$$\pm$$0.01$ \\ [1.0ex]
Q$1019$$+$$309$G1 & $0.3461  $ & $0.70 $$\pm$$0.07 $  & $ \cdots$        &$\cdots$& $ \cdots$          & $ \cdots$              & $ \cdots$                & $ \cdots$          & $ \cdots$         & $ \cdots$       \\ [1.0ex]
Q$1038$$+$$064$G1 & $0.441503$ & $0.673$$\pm$$0.011$  &$1.34$$\pm$$0.04$ &    7   & $13.73$$\pm$$0.04$ & 92.8$^{+12.2}_{-8.3}$  & 13.61$^{+ 0.04}_{-0.04}$ & $10.39$$\pm$$0.88$ & $40.3$$\pm$$0.7$  & $0.36$$\pm$$0.03$ \\ [1.0ex]
Q$1127$$-$$145$G1 & $0.328112$ & $0.028$$\pm$$0.003$  &$1.56$$\pm$$0.25$ &    2   & $11.92$$\pm$$0.03$ & 3.4$^{+0.5}_{-0.7}$    & 11.87$^{+ 0.07}_{-0.09}$ & $49.34$$\pm$$5.49$ & $61.3$$\pm$$2.9$  & $13.19$$\pm$$1.76$ \\ [1.0ex]
Q$1127$$-$$145$G2 & $0.328112$ & $0.028$$\pm$$0.003$  &$1.56$$\pm$$0.25$ &    2   & $11.92$$\pm$$0.03$ & 3.4$^{+0.5}_{-0.7}$    & 11.87$^{+ 0.07}_{-0.09}$ & $49.34$$\pm$$5.49$ & $61.3$$\pm$$2.9$  & $13.19$$\pm$$1.66$ \\ [1.0ex]
Q$1148$$+$$387$G1 & $0.553499$ & $0.640$$\pm$$0.013$  &$1.75$$\pm$$0.07$ &    8   & $13.48$$\pm$$0.02$ & 113.4$^{+8.3}_{-12.8}$ & 13.40$^{+ 0.03}_{-0.05}$ & $-2.69$$\pm$$1.24$ & $49.2$$\pm$$1.0$  & $0.24$$\pm$$0.17$ \\ [1.0ex]
Q$1209$$+$$107$G1 & $0.392857$ & $1.187$$\pm$$0.005$  &$1.44$$\pm$$0.01$ &   12   & $13.94$$\pm$$0.06$ & 258.5$^{+17.9}_{-20.8}$& 13.76$^{+ 0.03}_{-0.04}$ & $-40.39$$\pm$$0.49$& $95.9$$\pm$$0.4$  & $-0.66$$\pm$$0.01$ \\ [1.0ex]
Q$1222$$+$$228$G1 & $0.550188$ & $0.094$$\pm$$0.009$  &$1.56$$\pm$$0.24$ &    3   & $12.45$$\pm$$0.12$ & 11.0$^{+1.1}_{-2.1}$   & 12.38$^{+ 0.04}_{-0.08}$ & $-0.63$$\pm$$1.61$ & $13.0$$\pm$$1.5$  & $-0.98$$\pm$$0.34$ \\ [1.0ex]
Q$1241$$+$$176$G1 & $0.550493$ & $0.465$$\pm$$0.011$  &$1.29$$\pm$$0.04$ &    4   & $13.64$$\pm$$0.07$ & 95.8$^{+5.6}_{-8.2}$   & 13.32$^{+ 0.03}_{-0.04}$ & $10.18$$\pm$$1.53$ & $41.5$$\pm$$2.3$  & $1.07$$\pm$$0.06$ \\ [1.0ex]
Q$1246$$-$$057$G1 & $0.639913$ & $0.450$$\pm$$0.004$  &$1.18$$\pm$$0.02$ &    3   & $13.75$$\pm$$0.10$ & $>$101.3               & $>$13.65                 & $0.85$$\pm$$0.30$  & $15.9$$\pm$$0.4$  & $0.14$$\pm$$0.06$ \\ [1.0ex]
Q$1317$$+$$277$G1 & $0.660051$ & $0.320$$\pm$$0.006$  &$1.61$$\pm$$0.06$ &    6   & $13.14$$\pm$$0.07$ & 53.5$^{+5.8}_{-6.9}$   & 13.07$^{+ 0.05}_{-0.06}$ & $22.46$$\pm$$0.98$ & $41.4$$\pm$$0.9$  & $1.18$$\pm$$0.04$ \\ [1.0ex]
Q$1332$$+$$552$G1 & $0.374   $ & $2.90 $\phantom{000.000}& $ \cdots$     &$\cdots$& $ \cdots$          & $ \cdots$              & $ \cdots$                & $ \cdots$          & $ \cdots$         & $ \cdots$       \\ [1.0ex]
Q$1354$$+$$195$G1 & $0.456605$ & $0.773$$\pm$$0.015$  &$1.33$$\pm$$0.04$ &    6   & $13.91$$\pm$$0.38$ & 174.4$^{+10.2}_{-20.0}$& 13.58$^{+ 0.03}_{-0.05}$ & $-8.65$$\pm$$0.94$ & $35.1$$\pm$$0.9$  & $-0.35$$\pm$$0.02$ \\ [1.0ex]
Q$1424$$-$$118$G1 & $0.341712$ & $0.100$$\pm$$0.015$  &$1.87$$\pm$$0.39$ &    1   & $12.61$$\pm$$0.06$ & 20.2$^{+2.0}_{-7.3}$   & 12.65$^{+ 0.04}_{-0.16}$ & $0.09$$\pm$$1.033$ & $3.7$$\pm$$1.3$   & $-0.26$$\pm$$0.37$ \\ [1.0ex]
Q$1511$$+$$103$G1 & $0.4369  $ & $0.454$$\pm$$0.046$  &$ \cdots$         &$\cdots$& $ \cdots$          & $ \cdots$              & $ \cdots$                & $ \cdots$          & $ \cdots$         & $ \cdots$       \\ [1.0ex]
Q$1622$$+$$235$G1 & $0.317712$ & $0.491$$\pm$$0.010$  &$1.23$$\pm$$0.04$ &    6   & $13.88$$\pm$$0.58$ & 112.0$^{+6.6}_{-8.9}$  & 13.39$^{+ 0.03}_{-0.04}$ & $26.06$$\pm$$0.98$ & $41.4$$\pm$$0.8$  & $0.64$$\pm$$0.02$ \\ [1.0ex]
Q$1622$$+$$235$G2 & $0.368106$ & $0.247$$\pm$$0.005$  &$1.25$$\pm$$0.05$ &    3   & $13.21$$\pm$$1.13$ & 45.7$^{+5.4}_{-6.7}$   & 13.00$^{+ 0.05}_{-0.06}$ & $-2.03$$\pm$$0.42$ & $11.4$$\pm$$0.4$  & $-0.27$$\pm$$0.06$ \\ [1.0ex]
Q$1622$$+$$235$G3 & $0.471928$ & $0.769$$\pm$$0.006$  &$1.19$$\pm$$0.02$ &    8   & $14.04$$\pm$$0.20$ & 147.4$^{+14.4}_{-11.2}$& 13.81$^{+ 0.04}_{-0.03}$ & $-0.32$$\pm$$0.38$ & $28.1$$\pm$$0.4$  & $-0.16$$\pm$$0.03$ \\ [1.0ex]
Q$1622$$+$$235$G4 & $0.656069$ & $1.446$$\pm$$0.006$  &$1.10$$\pm$$0.01$ &   17   & $15.18$$\pm$$0.06$ & 545.6$^{+12.4}_{-17.5}$& 14.08$^{+ 0.01}_{-0.01}$ & $-6.42$$\pm$$0.34$ & $46.0$$\pm$$0.3$  & $-0.13$$\pm$$0.02$ \\[1.0ex] 
Q$1622$$+$$235$G5 & $0.702904$ & $0.032$$\pm$$0.003$  &$1.61$$\pm$$0.27$ &    1   & $12.09$$\pm$$0.05$ & 4.4$^{+0.9}_{-1.2}$    & 11.98$^{+ 0.09}_{-0.12}$ & $0.41$$\pm$$0.63$  & $2.4$$\pm$$1.1$   & $0.80$$\pm$$0.67$ \\ [1.0ex]
Q$1622$$+$$235$G6 & $0.797061$ & $0.468$$\pm$$0.008$  &$1.62$$\pm$$0.05$ &    6   & $13.28$$\pm$$0.01$ & 79.7$^{+7.1}_{-8.3}$   & 13.24$^{+ 0.04}_{-0.04}$ & $-9.92$$\pm$$1.24$ & $55.5$$\pm$$1.4$  & $-0.96$$\pm$$0.06$ \\ [1.0ex]
Q$1622$$+$$235$G7 & $0.891190$ & $1.548$$\pm$$0.004$  &$1.09$$\pm$$0.01$ &   10   & $15.07$$\pm$$0.02$ & $>$529.3               & $>$14.37                 & $-13.42$$\pm$$0.22$& $52.2$$\pm$$0.2$  & $-0.18$$\pm$$0.01$ \\ [1.0ex]
Q$1623$$+$$268$G1 & $0.887679$ & $0.903$$\pm$$0.004$  &$1.25$$\pm$$0.01$ &   10   & $14.58$$\pm$$0.03$ & 219.0$^{+20.8}_{-17.4}$& 13.68$^{+ 0.04}_{-0.04}$ & $5.42$$\pm$$0.35$  & $75.0$$\pm$$0.25$ & $0.45$$\pm$$0.01$  \\ [1.0ex]
Q$1623$$+$$268$G2 & $0.887679$ & $0.903$$\pm$$0.004$  &$1.25$$\pm$$0.01$ &   10   & $14.58$$\pm$$0.03$ & 219.0$^{+20.8}_{-17.4}$& 13.68$^{+ 0.04}_{-0.04}$ & $5.42$$\pm$$0.35$  & $75.0$$\pm$$0.25$ & $0.45$$\pm$$0.01$  \\ [1.0ex]
Q$2128$$-$$123$G1 & $0.429708$ & $0.395$$\pm$$0.010$  &$1.17$$\pm$$0.05$ &    5   & $14.06$$\pm$$0.36$ & 94.4$^{+13.4}_{-14.6}$ & 13.32$^{+ 0.06}_{-0.07}$ & $-5.45$$\pm$$0.91$ & $19.3$$\pm$$1.0$  & $-0.38$$\pm$$0.05$ \\ [1.0ex]
Q$2206$$-$$199$G1 & $0.948381$ & $0.249$$\pm$$0.002$  &$1.33$$\pm$$0.02$ &    5   & $13.25$$\pm$$0.07$ & 55.2$^{+8.1}_{-8.5}$   & 13.09$^{+ 0.06}_{-0.07}$ & $-23.84$$\pm$$0.35$& $41.3$$\pm$$0.2$  & $-1.45$$\pm$$0.01$ \\ [1.0ex]
Q$2206$$-$$199$G2 & $1.016979$ & $1.047$$\pm$$0.003$  &$1.14$$\pm$$0.01$ &   10   & $14.67$$\pm$$0.04$ & 405.2$^{+26.5}_{-32.5}$& 13.95$^{+ 0.03}_{-0.04}$ & $-2.13$$\pm$$0.34$ & $49.0$$\pm$$0.6$  & $0.53$$\pm$$0.01$ \\ \hline
\end{tabular}}
\end{center}
$^a$ In cases where the {\MgII} $\lambda2796$ absorption is saturated,
we use the {\MgII} $\lambda2803$ absorption profile to compute
log$(N_a)$ and $\tau$. The {\MgII} $\lambda2796$ and {\MgII}
$\lambda2803$ absorption profiles are both saturated in only six cases.
\end{table*}

\begin{table*}
\begin{center}
  \caption{The {\MgII} absorbing galaxy morphological properties,
which were obtained from GIM2D modeling of the WFPC-2/{\it HST}
images. The table columns are (1) the quasar name and galaxy ID
number, (2) the optical depth weighted mean {\MgII} $\lambda2796$
absorption redshift ($z_{abs}$), (3) the {\MgII} $\lambda2796$
rest-frame equivalent width $W_r(2796)$, (4) the bulge-to-total
fraction ($B/T$), (5) the bulge S{\'e}rsic index ($n$) (6) the bulge
semi-major axis effective radius ($r_b$), (7) the bulge ellipticity
($e_b\equiv 1-b/a$), (8) the bulge position angle ($\theta_b$), (9)
the semi-major axis disk scale length ($r_d$), (10) the disk
inclination ($i$), (11) the disk position angle ($\theta_d$), (12) the
galaxy half light radius ($r_h$), and (13) the $\chi^2$ per degree of
freedom for the fit.}
  \vspace{-0.5em}
\label{tab:galmorph2}
{\footnotesize\begin{tabular}{llcccccrcrccc}\hline
QSO Field         & \phantom{00}$z_{abs}$    & $W_r(2796)$ & $B/T$                    & $n$                   & $r_b$                 & $       e_b$             & $\theta _b$        & $r_d$\phantom{000}    & $i$                    & $\theta _d$\phantom{0}   & $r_{h}$            & $\chi ^2$\\
                  &                          &    (\AA)    &                          &                       &  (kpc)                &                          &   (deg)            & (kpc)\phantom{0}      &  (deg)                 &  (deg)                  & (kpc)                    &          \\\hline
Q$ 0002$$+$$051$G1 & $ 0.298059 $ &$ 0.244$$\pm$$0.003 $   & $ 0.41_{-0.07}^{+0.06} $ & $ 2.9_{-0.4}^{+0.5} $ & $ 1.3_{-0.3}^{+0.3} $ & $ 0.13_{-0.03}^{+0.05} $ & $ 65_{-13}^{+11} $ & $ 3.2_{-0.1}^{+0.1} $ & $ 29.5_{-4.5}^{+2.4} $ &$ 50_{-8}^{+12} $  & $ 3.47 $ & $ 0.97 $ \\[1.0ex]
Q$ 0002$$+$$051$G2 & $ 0.591365 $ &$ 0.102$$\pm$$0.002 $   & $ 0.41_{-0.05}^{+0.06} $ & $ 1.8_{-0.7}^{+0.6} $ & $ 0.9_{-0.1}^{+0.2} $ & $ 0.19_{-0.09}^{+0.09} $ & $ 18_{-13}^{+12} $ & $ 4.1_{-0.6}^{+0.6} $ & $ 40.7_{-3.7}^{+4.3} $ &$ 77_{-10}^{+10} $ & $ 3.35 $ & $ 0.97 $ \\[1.0ex]
Q$ 0002$$+$$051$G3 & $ 0.851393 $ &$ 1.089$$\pm$$0.008 $   & $ 0.45_{-0.04}^{+0.06} $ & $ 0.5_{-0.3}^{+0.5} $ & $ 1.1_{-0.1}^{+0.1} $ & $ 0.55_{-0.09}^{+0.09} $ & $ 20_{-5}^{+4} $   & $ 4.0_{-0.5}^{+0.5} $ & $ 48.1_{-10.5}^{+6.7} $&$ 7_{-12}^{+12} $  & $ 2.47 $ & $ 0.87 $ \\[1.0ex]
Q$ 0109$$+$$200$G1 & $ 0.5346   $ &$2.26$\phantom{0000008} & $ 0.52_{-0.12}^{+0.14} $ & $ 1.4_{-0.4}^{+0.5} $ & $ 1.3_{-0.2}^{+0.2} $ & $ 0.26_{-0.12}^{+0.26} $ & $ 34_{-17}^{+14} $ & $ 2.5_{-0.6}^{+0.4} $ & $ 65.2_{-7.0}^{+6.4} $ &$ 47_{-8}^{+13} $  & $ 2.24 $ & $ 0.92 $ \\[1.0ex]
Q$ 0117$$+$$213$G1 & $ 0.729077 $ & $ 0.244$$\pm$$0.005 $  & $ 0.24_{-0.09}^{+0.17} $ & $ 3.7_{-1.4}^{+0.3} $ & $ 1.0_{-0.7}^{+1.1} $ & $ 0.46_{-0.10}^{+0.14} $ & $ 90_{-23}^{+13} $ & $ 3.3_{-0.5}^{+0.9} $ & $ 45.4_{-7.0}^{+12.1} $&$ 51_{-10}^{+11} $ & $ 4.41 $ & $ 0.97 $ \\[1.0ex]
Q$ 0150$$-$$202$G1 & $ 0.3887   $ & $ 0.580$$\pm$$0.050 $  & $ 0.13_{-0.03}^{+0.02} $ & $ 3.5_{-0.2}^{+0.2} $ & $ 2.1_{-0.5}^{+0.4} $ & $ 0.39_{-0.08}^{+0.08} $ & $ 88_{-13}^{+16} $ & $ 4.6_{-0.2}^{+0.2} $ & $ 76.4_{-1.0}^{+0.7} $ &$ 81_{-1}^{+1} $   & $ 7.03 $ & $ 0.99 $ \\[1.0ex]
Q$ 0229$$+$$131$G1 & $ 0.417338 $ & $ 0.816$$\pm$$0.020 $  & $ 0.29_{-0.01}^{+0.01} $ & $ 1.2_{-0.1}^{+0.1} $ & $ 1.2_{-0.0}^{+0.0} $ & $ 0.17_{-0.03}^{+0.03} $ & $ 38_{-6}^{+6} $   & $ 6.7_{-0.2}^{+0.2} $ & $ 51.4_{-1.3}^{+1.0} $ &$ 82_{-1}^{+1} $   & $ 7.36 $ & $ 1.37 $ \\[1.0ex]
Q$ 0235$$+$$164$G1 & $ 0.524    $ & $ 2.34 $$\pm$$0.05  $  & $ 0.33_{-0.02}^{+0.03} $ & $ 1.9_{-1.7}^{+2.1} $ & $ 0.1_{-0.1}^{+0.1} $ & $ 0.59_{-0.19}^{+0.11} $ & $ 54_{-102}^{+63} $& $ 3.2_{-0.4}^{+0.4} $ & $ 42.0_{-7.9}^{+5.4} $ &$ 10_{-13}^{+14} $ & $ 3.10 $ & $ 1.19 $ \\[1.0ex]
Q$ 0349$$-$$146$G1 & $ 0.357168 $ & $ 0.175\pm0.007 $      & $ 0.27_{-0.04}^{+0.08} $ & $ 0.3_{-0.1}^{+0.1} $ & $ 3.1_{-0.2}^{+0.1} $ & $ 0.46_{-0.05}^{+0.06} $ & $ 62_{-3}^{+4} $   & $ 2.3_{-0.1}^{+0.2} $ & $ 56.3_{-4.0}^{+-5.6} $&$ 77_{-7}^{+8} $   & $ 3.66 $ & $ 1.16 $ \\[1.0ex]
Q$ 0450$$-$$132$G1 & $ 0.493936 $ & $ 0.674\pm0.024 $      & $ 0.55_{-0.05}^{+0.05} $ & $ 0.8_{-0.1}^{+0.1} $ & $ 1.7_{-0.1}^{+0.1} $ & $ 0.55_{-0.03}^{+0.03} $ & $ 22_{-2}^{+3} $   & $ 3.5_{-0.4}^{+0.2} $ & $ 60.0_{-1.1}^{+0.0} $ &$ 62_{-3}^{+4} $   & $ 2.68 $ & $ 3.45 $ \\[1.0ex]
Q$ 0450$$-$$132$G2 & $ 0.493936 $ &$ 0.674$$\pm$$0.024 $   & $ 0.62_{-0.03}^{+0.06} $ & $ 0.7_{-0.1}^{+0.2} $ & $ 1.8_{-0.1}^{+0.1} $ & $ 0.70_{-0.01}^{+0.00} $ & $ 11_{-2}^{+2} $   & $ 2.6_{-0.2}^{+0.1} $ & $ 63.6_{-3.6}^{+3.0} $ &$ 23_{-4}^{+6} $   & $ 2.32 $ & $ 1.63 $ \\[1.0ex]
Q$ 0454$$-$$220$G1 & $ 0.483337 $ & $ 0.426\pm0.007 $      & $ 0.00_{-0.00}^{+0.10} $ & $ 3.9_{-0.1}^{+0.1} $ & $ 4.2_{-0.1}^{+0.4} $ & $ 0.37_{-0.01}^{+0.01} $ & $ 2_{-7}^{+1} $    & $ 4.9_{-0.2}^{+0.1} $ & $ 42.1_{-3.1}^{+2.7} $ &$ 5_{-4}^{+4} $    & $ 8.26 $ & $ 1.16 $ \\[1.0ex]
Q$ 0827$$+$$243$G1 & $ 0.524966 $ & $ 2.419$$\pm$$0.012 $  & $ 0.00_{-0.00}^{+0.01} $ & $ 4.0_{-0.1}^{+0.0} $ & $ 2.9_{-0.5}^{+1.1} $ & $ 0.32_{-0.02}^{+0.02} $ & $ 77_{-3}^{+8} $   & $ 4.6_{-0.1}^{+0.1} $ & $ 74.4_{-0.7}^{+0.6} $ &$ 87_{-1}^{+1} $   & $ 7.64 $ & $ 1.38 $ \\[1.0ex]
Q$ 0836$$+$$113$G1 & $ 0.786725 $ & $ 2.133$$\pm$$0.019 $  & $ 0.48_{-0.08}^{+0.09} $ & $ 0.3_{-0.1}^{+0.2} $ & $ 4.6_{-0.3}^{+0.4} $ & $ 0.70_{-0.04}^{+0.00} $ & $ 17_{-3}^{+4} $   & $ 2.3_{-0.1}^{+0.2} $ & $ 84.9_{-1.1}^{+0.1} $ &$ 32_{-1}^{+2} $   & $ 4.36 $ & $ 1.31 $ \\[1.0ex]
Q$ 1019$$+$$309$G1 & $ 0.3461   $ & $ 0.70 $$\pm$$0.07  $  & $ 0.33_{-0.04}^{+0.04} $ & $ 0.2_{-0.0}^{+0.1} $ & $ 2.3_{-0.1}^{+0.1} $ & $ 0.51_{-0.02}^{+0.03} $ & $ 6_{-2}^{+2} $    & $ 1.9_{-0.1}^{+0.1} $ & $ 38.6_{-3.4}^{+4.5} $ &$ 10_{-7}^{+7} $   & $ 2.91 $ & $ 1.07 $ \\[1.0ex]
Q$ 1038$$+$$064$G1 & $ 0.441453 $ & $ 0.673$$\pm$$0.011 $  & $ 0.73_{-0.09}^{+0.08} $ & $ 0.6_{-0.1}^{+0.1} $ & $ 4.6_{-0.1}^{+0.1} $ & $ 0.64_{-0.02}^{+0.01} $ & $ 86_{-1}^{+1} $   & $ 4.1_{-0.3}^{+0.3} $ & $ 49.8_{-5.2}^{+7.4} $ &$ 78_{-4}^{+4} $   & $ 5.04 $ & $ 1.07 $ \\[1.0ex]
Q$ 1127$$-$$145$G1 & $ 0.328279 $ & $ 0.028$$\pm$$0.003 $  & $ 0.40_{-0.07}^{+0.09} $ & $ 2.6_{-0.4}^{+0.7} $ & $ 0.7_{-0.1}^{+0.2} $ & $ 0.21_{-0.06}^{+0.07} $ & $ 84_{-13}^{+13} $ & $ 2.1_{-0.2}^{+0.2} $ & $ 37.6_{-4.4}^{+3.1} $ &$ 73_{-7}^{+8} $   & $ 2.10 $ & $ 0.97 $ \\[1.0ex]
Q$ 1127$$-$$145$G2 & $ 0.328279 $ & $ 0.028$$\pm$$0.003 $  & $ 0.05_{-0.00}^{+0.00} $ & $ 0.5_{-0.3}^{+0.5} $ & $ 0.4_{-0.1}^{+0.1} $ & $ 0.07_{-0.04}^{+0.05} $ & $ 70_{-18}^{+13} $ & $ 2.7_{-0.0}^{+0.1} $ & $ 22.1_{-2.7}^{+4.0} $ &$ 30_{-7}^{+7} $   & $ 4.27 $ & $ 1.06 $ \\[1.0ex]
Q$ 1148$$+$$387$G1 & $ 0.553363 $ & $ 0.640\pm0.013 $      & $ 0.07_{-0.06}^{+0.02} $ & $ 2.9_{-0.8}^{+0.6} $ & $ 3.3_{-3.3}^{+1.1} $ & $ 0.06_{-0.06}^{+0.05} $ & $ 11_{-48}^{+17} $ & $ 4.1_{-0.2}^{+0.2} $ & $ 49.8_{-3.0}^{+2.2} $ &$ 28_{-2}^{+2} $   & $ 6.76 $ & $ 1.10 $ \\[1.0ex]
Q$ 1209$$+$$107$G1 & $ 0.392924 $ & $ 1.187$$\pm$$0.005 $  & $ 0.61_{-0.08}^{+0.10} $ & $ 0.6_{-0.1}^{+0.1} $ & $ 0.9_{-0.0}^{+0.0} $ & $ 0.60_{-0.05}^{+0.04} $ & $ 2_{-3}^{+3} $    & $ 1.2_{-0.3}^{+0.1} $ & $ 47.2_{-5.1}^{+6.1} $ &$ 14_{-19}^{+11} $ & $ 1.21 $ & $ 1.21 $ \\[1.0ex]
Q$ 1222$$+$$228$G1 & $ 0.550198 $ & $ 0.094$$\pm$$0.009 $  & $ 0.00_{-0.00}^{+0.03} $ & $ 2.0_{-0.6}^{+0.2} $ & $ 1.9_{-0.6}^{+0.3} $ & $ 0.32_{-0.03}^{+0.05} $ & $ 48_{-18}^{+13} $ & $ 3.3_{-0.2}^{+0.2} $ & $ 81.3_{-1.0}^{+1.0} $ &$ 78_{-1}^{+1} $   & $ 5.51 $ & $ 1.01 $ \\[1.0ex]
Q$ 1241$$+$$176$G1 & $ 0.550482 $ & $ 0.465$$\pm$$0.011 $  & $ 0.55_{-0.10}^{+0.12} $ & $ 3.7_{-0.6}^{+0.3} $ & $ 1.5_{-0.4}^{+0.7} $ & $ 0.45_{-0.05}^{+0.07} $ & $ 33_{-5}^{+5} $   & $ 2.1_{-0.2}^{+0.1} $ & $ 31.7_{-4.8}^{+16.2} $&$ 70_{-19}^{+17} $ & $ 2.49 $ & $ 1.21 $ \\[1.0ex]
Q$ 1246$$-$$057$G1 & $ 0.639909 $ & $ 0.450\pm0.004 $      & $ 0.31_{-0.15}^{+0.22} $ & $ 0.2_{-0.0}^{+0.2} $ & $ 5.1_{-0.8}^{+0.8} $ & $ 0.36_{-0.22}^{+0.13} $ & $ 20_{-26}^{+13} $ & $ 2.7_{-0.8}^{+0.4} $ & $ 28.6_{-11.7}^{+10.5}$&$ 48_{-25}^{+40} $ & $ 4.90 $ & $ 1.23 $ \\[1.0ex]
Q$ 1317$$+$$277$G1 & $ 0.660049 $ & $ 0.320$$\pm$$0.006 $  & $ 0.19_{-0.02}^{+0.02} $ & $ 0.9_{-0.4}^{+0.4} $ & $ 0.7_{-0.1}^{+0.1} $ & $ 0.46_{-0.07}^{+0.10} $ & $ 12_{-10}^{+8} $  & $ 3.9_{-0.2}^{+0.2} $ & $ 65.8_{-1.2}^{+1.2} $ &$ 3_{-1}^{+1} $    & $ 5.11 $ & $ 1.12 $ \\[1.0ex]
Q$ 1332$$+$$552$G1 & $ 0.374    $ &$ 2.90$\phantom{0000008}& $ 0.79_{-0.03}^{+0.03}$  & $ 3.8_{-0.2}^{+0.2} $ & $ 3.9_{-0.2}^{+0.3} $ & $ 0.19_{-0.02}^{+0.03} $ & $ 30_{-4}^{+5} $   & $ 4.2_{-0.3}^{+0.3} $ & $ 55.6_{-3.0}^{+2.5} $ &$ 83_{-4}^{+4} $   & $ 4.71 $ & $ 1.06 $ \\[1.0ex]
Q$ 1354$$+$$195$G1 & $ 0.456598 $ & $ 0.773\pm0.015 $      & $ 0.00_{-0.00}^{+0.02} $ & $ 2.7_{-0.4}^{+0.3} $ & $ 5.1_{-1.8}^{+1.9} $ & $ 0.37_{-0.10}^{+0.14} $ & $ 27_{-10}^{+13} $ & $ 2.4_{-0.1}^{+0.1} $ & $ 24.7_{-6.5}^{+5.7} $ &$ 26_{-14}^{+14} $ & $ 4.04 $ & $ 1.06 $ \\[1.0ex]
Q$ 1424$$-$$118$G1 & $ 0.341716 $ & $ 0.100\pm0.015 $      & $ 0.00_{-0.00}^{+0.01} $ & $ 3.8_{-0.0}^{+0.1} $ & $ 2.4_{-0.3}^{+0.1} $ & $ 0.57_{-0.02}^{+0.03} $ & $ 76_{-2}^{+1} $   & $ 4.6_{-0.1}^{+0.1} $ & $ 66.2_{-0.4}^{+0.5} $ &$ 89_{-1}^{+1} $   & $ 7.72 $ & $ 1.14 $ \\[1.0ex]
Q$ 1511$$+$$103$G1 & $ 0.4369   $ &$ 0.454$$\pm$$0.046 $   & $ 0.00_{-0.00}^{+0.04} $ & $ 0.7_{-0.2}^{+0.1} $ & $ 1.9_{-0.3}^{+1.0} $ & $ 0.33_{-0.04}^{+0.03} $ & $ 79_{-27}^{+10} $ & $ 2.3_{-0.1}^{+0.1} $ & $ 51.0_{-2.2}^{+1.8} $ &$ 76_{-2}^{+2} $   & $ 3.79 $ & $ 1.06 $ \\[1.0ex]
Q$ 1622$$+$$235$G1 & $ 0.317597 $ &$ 0.491$$\pm$$0.010 $   & $ 0.64_{-0.02}^{+0.02} $ & $ 3.1_{-0.1}^{+0.1} $ & $ 1.0_{-0.0}^{+0.1} $ & $ 0.31_{-0.02}^{+0.03} $ & $ 44_{-2}^{+2} $   & $ 4.9_{-0.2}^{+0.1} $ & $ 70.6_{-0.8}^{+1.0} $ &$ 25_{-1}^{+1} $   & $ 2.20 $ & $ 0.95 $ \\[1.0ex]
Q$ 1622$$+$$235$G2 & $ 0.368112 $ &$ 0.247$$\pm$$0.005 $   & $ 0.39_{-0.01}^{+0.02} $ & $ 4.0_{-0.0}^{+0.0} $ & $ 3.4_{-0.1}^{+0.2} $ & $ 0.38_{-0.04}^{+0.02} $ & $ 35_{-3}^{+3} $   & $ 3.4_{-0.1}^{+0.0} $ & $ 64.2_{-0.7}^{+0.7} $ &$ 5_{-1}^{+1} $    & $ 5.03 $ & $ 1.07 $ \\[1.0ex]
Q$ 1622$$+$$235$G3 & $ 0.471930 $ &$ 0.769$$\pm$$0.006 $   & $ 0.00_{-0.00}^{+0.02} $ & $ 1.9_{-0.1}^{+0.1} $ & $ 0.9_{-0.2}^{+0.1} $ & $ 0.32_{-0.02}^{+0.02} $ & $ 31_{-23}^{+14} $ & $ 1.2_{-0.0}^{+0.0} $ & $ 71.4_{-1.2}^{+1.2} $ &$ 6_{-1}^{+1} $    & $ 2.02 $ & $ 1.07 $ \\[1.0ex]
Q$ 1622$$+$$235$G4 & $ 0.656106 $ &$ 1.446$$\pm$$0.006 $   & $ 0.25_{-0.10}^{+0.09} $ & $ 0.2_{-0.0}^{+0.3} $ & $ 3.1_{-0.4}^{+0.3} $ & $ 0.18_{-0.09}^{+0.11} $ & $ 41_{-15}^{+16} $ & $ 3.8_{-0.4}^{+0.4} $ & $ 65.6_{-3.4}^{+2.4} $ &$ 3_{-3}^{+4} $    & $ 4.95 $ & $ 1.06 $ \\[1.0ex]
Q$ 1622$$+$$235$G5 & $ 0.702902 $ &$ 0.032$$\pm$$0.003 $   & $ 0.03_{-0.02}^{+0.01} $ & $ 0.4_{-0.2}^{+0.1} $ & $ 1.2_{-0.4}^{+0.4} $ & $ 0.69_{-0.05}^{+0.01} $ & $ 82_{-18}^{+16} $ & $ 2.9_{-0.1}^{+0.1} $ & $ 41.7_{-3.2}^{+1.5} $ &$ 29_{-3}^{+3} $   & $ 4.69 $ & $ 1.08 $ \\[1.0ex]
Q$ 1622$$+$$235$G6 & $ 0.797078 $ &$ 0.468$$\pm$$0.008 $   & $ 0.95_{-0.09}^{+0.05} $ & $ 0.5_{-0.1}^{+0.0} $ & $ 5.8_{-0.2}^{+0.2} $ & $ 0.58_{-0.02}^{+0.02} $ & $ 29_{-2}^{+2} $   & $ 5.1_{-0.8}^{+0.6} $ & $ 77.9_{-6.2}^{+3.6} $ &$ 21_{-9}^{+5} $   & $ 5.99 $ & $ 0.95 $ \\[1.0ex]
Q$ 1622$$+$$235$G7 & $ 0.891276 $ &$ 1.548$$\pm$$0.004 $   & $ 0.04_{-0.02}^{+0.02} $ & $ 1.4_{-0.4}^{+0.3} $ & $ 0.5_{-0.3}^{+0.4} $ & $ 0.66_{-0.04}^{+0.04} $ & $ 27_{-13}^{+18} $ & $ 4.4_{-0.3}^{+0.3} $ & $ 71.5_{-1.7}^{+1.4} $ &$ 35_{-1}^{+2} $   & $ 7.19 $ & $ 1.15 $ \\[1.0ex]
Q$ 1623$$+$$268$G1 & $ 0.887679 $ &$ 0.903$$\pm$$0.004 $   & $ 0.55_{-0.55}^{+0.15} $ & $ 0.4_{-0.2}^{+0.8} $ & $ 7.5_{-3.7}^{+1.7} $ & $ 0.52_{-0.51}^{+0.17} $ & $ 73_{-24}^{+11} $ & $ 3.6_{-1.4}^{+0.9} $ & $ 75.1_{-16.3}^{+7.7} $&$ 64_{-5}^{+17} $  & $ 7.00 $ & $ 1.05 $ \\[1.0ex]
Q$ 1623$$+$$268$G2 & $ 0.887679 $ &$ 0.903$$\pm$$0.004 $   & $ 0.74_{-0.21}^{+0.16} $ & $ 3.7_{-0.8}^{+0.3} $ & $ 2.0_{-0.3}^{+0.3} $ & $ 0.57_{-0.18}^{+0.13} $ & $ 64_{-12}^{+18} $ & $ 1.5_{-0.4}^{+0.7} $ & $ 74.2_{-12.8}^{+10.8}$&$ 78_{-12}^{+11} $ & $ 2.25 $ & $ 1.10 $ \\[1.0ex]
Q$ 2128$$-$$123$G1 & $ 0.429735 $ &$ 0.395$$\pm$$0.010 $   & $ 0.07_{-0.03}^{+0.04} $ & $ 3.5_{-0.7}^{+0.5} $ & $ 1.6_{-0.7}^{+1.5} $ & $ 0.56_{-0.12}^{+0.14} $ & $ 33_{-19}^{+19} $ & $ 2.9_{-0.2}^{+0.2} $ & $ 48.3_{-3.7}^{+3.5} $ &$ 75_{-5}^{+6} $   & $ 4.65 $ & $ 0.96 $ \\[1.0ex]
Q$ 2206$$-$$199$G1 & $ 0.948361 $ & $ 0.249\pm0.002 $      & $ 0.64_{-0.13}^{+0.10} $ & $ 3.1_{-0.5}^{+0.4} $ & $ 1.9_{-0.2}^{+0.2} $ & $ 0.47_{-0.09}^{+0.07} $ & $ 11_{-9}^{+7} $   & $ 2.2_{-0.3}^{+0.4} $ & $ 73.2_{-4.5}^{+3.9} $ &$ 23_{-4}^{+7} $   & $ 2.59 $ & $ 1.01 $ \\[1.0ex]
Q$ 2206$$-$$199$G2 & $ 1.017038 $ & $ 1.047\pm0.003 $      & $ 1.00_{-0.05}^{+0.00} $ & $ 0.5_{-0.0}^{+0.0} $ & $ 9.5_{-0.4}^{+0.4} $ & $ 0.07_{-0.04}^{+0.04} $ & $ 82_{-18}^{+12} $ & $ 22.3_{-1.8}^{+1.8} $ & $ 2.9_{-2.9}^{+17.4} $&$ 70_{-55}^{+20} $ & $ 9.62 $ & $ 1.65 $ \\[1.0ex] \hline
\end{tabular}}
\end{center}
\end{table*}

\section{Results}
\label{sec:results}

\begin{figure}
\includegraphics[angle=0,scale=0.87]{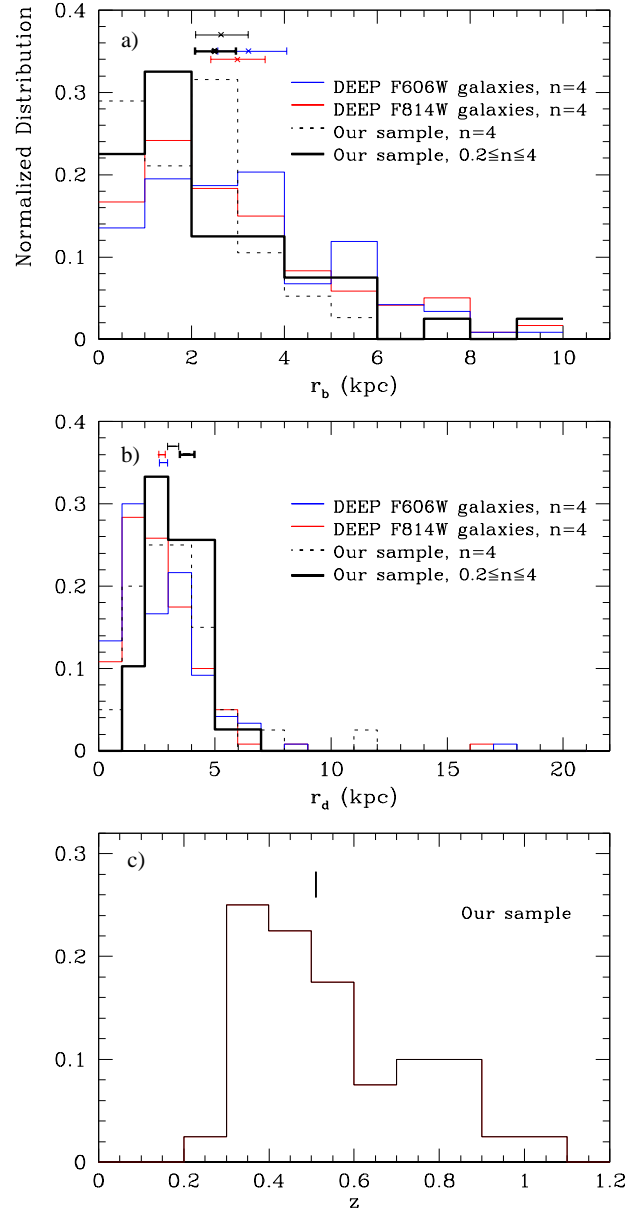}
\caption{--- (a) The normalized distributions of the bulge effective
radii derived for 120 galaxies from DEEP survey imaged with the F606W
filter using a fixed S{\'e}rsic index of $n=4$ (blue), 120 galaxies
from DEEP survey imaged with the F814W filter also using $n=4$ (red),
our sample having free S{\'e}rsic index fits (solid black), and our
sample fit with $n=4$ (dotted black). The mean values are also shown
and the error bars represent the average uncertainty from the model
fits of all galaxies. All samples have similar bulge effective radii
distributions.  --- (b) same as (a) except plotted for the disk scale
length distributions. Samples have similar distributions except our
sample has larger disk scale lengths. Note that when the S{\'e}rsic
index is fixed at $n=4$, the disk scale length distribution shifts
towards smaller sizes and becomes more consistent with the DEEP
sample. --- (c) the redshift distributions for the 40 galaxies from
our sample. The mean redshift value is indicated. The 120 DEEP
galaxies have been selected to have the same redshift distribution as
our sample.}
\label{fig:disk}
\end{figure}

\subsection{The Sample}

We have constructed a sample of 40 {\MgII} absorption-selected
galaxies between 0.3$<$$z$$<$1.0. In Figure~\ref{fig:mos} we show
WFPC-2 portraits of the galaxies. Note that the galaxy images
are 10 times larger than the $1.5\sigma$ galaxy isophotal area, giving
the galaxies the appearance of all being of similar size; the angular
and physical scales of each galaxy image is shown.  The encircled
arrow next to the portrait provides the direction to the quasar
relative to the galaxy. In Table~\ref{tab:galmorph} we list the
observed galaxy properties for the sample, which were mostly derived
from the WFPC-2 images. The galaxies range between $-18.6 \leq M_B
\leq -23.8$ and have a impact parameter range of $12 \leq D \leq
115$~kpc.

Also in Figure~\ref{fig:mos} we show the fitted {\MgII} $\lambda$2796
absorption profiles. The number of clouds is indicated by the
tickmarks above the fit. We do not have in hand the high resolution
spectra of six {\MgII} systems so we can only use their equivalent
widths in our analysis. Note the wide variety of {\MgII} $\lambda$2796
profile velocity widths and structures. The {\MgII} rest-frame
equivalent widths range between $0.03\leq W_r(2796) \leq 2.9$~\AA. In
Table~\ref{tab:abs} we present the {\MgII} absorption properties
measured from the profiles and obtained from the Voigt profiles fits.

Along with the {\it HST} images shown in Figure~\ref{fig:mos}, we also
include the galaxy GIM2D model and residual images. The model and
residual image is displayed using the same dynamic range as the WFPC-2
image. As mentioned above, the smooth GIM2D models do not include
degrees of freedom to fit spiral arms, {\HII} regions etc.~seen in
some galaxy images. For example, one can clearly see these structures
in the residual images of Q0229$+$131G1 (spiral arms) and
Q0454$-$220G1 ({\HII} regions).

In Table~\ref{tab:galmorph2} we list the galaxy morphological
properties extracted from the GIM2D models. The 40 galaxies have a
mean disk scale length of 3.8~kpc, which is larger than the estimated
value for the Milky Way of $2.3\pm0.6$~kpc \citep{hammer07}. The
sample mean bulge scale length is 2.5~kpc, which is comparable to
estimates of the Milky Way bulge of 2~kpc \citep{rich98}. The average
S{\'e}rsic index for the sample is $\left<n\right>$=1.9. We find 23
galaxies having well-modeled disk-like bulges with S{\'e}rsic indices
of $n<2$ and 17 galaxies having steeper bulge profiles with $n\geq2$.

\subsubsection{{\MgII} Absorbers Compared to DEEP Galaxies}

In order to determine if {\MgII} absorption-selected galaxies differ
from field galaxies, we compare the bulge and disk size distribution
from our sample to galaxies having similar properties obtained from
the Deep Extragalactic Exploratory Probe (DEEP) Groth Strip Galaxy
Redshift Survey
\citep[see][]{simard02,vogt05,weiner05}. \cite{simard02} used GIM2D to
model WFPC-2/{\it HST} F606W and 814W images of 7450 galaxies from the
DEEP survey. We selected a subset of 429 galaxies that have both a
redshift and magnitude range similar to those represented by our
sample (see Table~\ref{tab:galmorph}). The mean redshift for the 429
DEEP galaxies is $\left<z\right>=0.68$, peaking at $z\sim 0.9$. In
Figure~\ref{fig:disk}c we show the redshift distribution for our
sample having a mean redshift of $\left<z\right>=0.51$, peaking at
$z\sim0.4$. Since the redshift distributions for the two samples are
quite different, we used our redshift distribution as a selection
function to randomly select a subset of the 429 DEEP galaxies to make
a fair comparison. We randomly selected 120 DEEP galaxies that
reproduce the redshift distribution of our sample seen in
Figure~\ref{fig:disk}c. We produced several DEEP sub-samples all
resulting in the similar distributions of bulge effective radii and
disk scale lengths.

The galaxies from the DEEP sample were fit using GIM2D where the bulge
S{\'e}rsic index is limited to $n=4$, whereas we allowed the bulge
S{\'e}rsic index to vary between $0.2\leq n\leq 4.0$ for our galaxy
model fits. In order to make a direct comparison with the DEEP
galaxies, we refit our 40 galaxies with a fixed S{\'e}rsic index of
$n=4$. It is also worth noting that the majority of our galaxies were
observed using the WFPC-2 F702W filter, whereas the DEEP survey used
two different WFPC-2 filters; a bluer F606W filter and a redder F814W
filter. The model fits to the DEEP F606W and F814W images produce
similar size distributions (see Fig~\ref{fig:disk}a,b). Given that our
galaxy images were taken primarily using a filter with a central
wavelength between the central wavelength of the two filters used in
DEEP, we do not expect that the difference in filter band-pass used
will produce any shifts/differences in the disk and bulge size
distributions.

In Figure~\ref{fig:disk}a, we show the normalized distribution of
bulge effective radii averaged over several random sub-samples of 120
galaxies derived from the DEEP F606W images, the DEEP F814W images,
our sample of 40 galaxies having free S{\'e}rsic index fit, and our
sample fit with $n=4$.  The mean values along with the mean fit errors
are shown. For our sample, the mean value is consistent with those
derived from the DEEP sample, although more galaxies are found with
smaller bulge effective radii in the latter.  The K-S probability that
the DEEP sample and our sample (fit with $n=4$) are drawn from the
same population is ${\rm P(KS)} = 0.27$, which suggests that the two
distributions are similar and only differ at 85\%
c.l. ($1.11\sigma$).  Note that for our sample, when the S{\'e}rsic
index is fixed at $n=4$, the bulge sizes tend to increase on
average. As the bulge contribution increases, it is reflected as a
smaller size distribution of the galaxy disks as seen in
Figure~\ref{fig:disk}b.

In Figure~\ref{fig:disk}b, we show the disk scale length distribution.
Here we find that the averaged sub-samples of 120 DEEP galaxies have a
lower mean $r_d$ than found for our sample. On average, our sample has
more galaxies with larger $r_d$ compared to those from the DEEP
survey.  The K-S probability that the DEEP sample and our sample (fit
with $n=4$) are drawn from the same distribution is ${\rm P(KS)}
=0.27$, which suggests that the two distributions are similar and only
differ at the 73\% c.l. ($1.11\sigma$).  However, if we compare our
sample, when using a free S{\'e}rsic index fit, this increases to
$3.7\sigma$ and the mean value is now inconsistent with those of the
DEEP survey.

We find our sample, when fit with $n=4$, similar to the galaxies in
the DEEP sample: differences arise only when we have a free index
fit. This difference is only due to the fitting and not a physical
difference.

\subsection{Galaxy Type and Absorption}

It is difficult to classify the morphological type (i.e., E4, Sb, etc)
of a galaxy using only the quantified morphological parameters
presented here.  While early-type galaxies are expected to have high
S{\'e}rsic indices of $n\geq 2$, over 25\% of spirals have bulges with
$n>2$ \citep[see][]{weinzirl09}. Thus, the S{\'e}rsic index is not
ideal for classifying galaxy morphology. Instead we have used the
bulge-to-total ratio to separate early-type ($B/T>0.5$) and late-type
galaxies ($B/T<0.5$). We expect less than 8\% ($\pm 3$ galaxies)
contamination by applying a $B/T=0.5$ cutoff \citep{weinzirl09}.

\begin{figure}
\includegraphics[angle=0,scale=0.45]{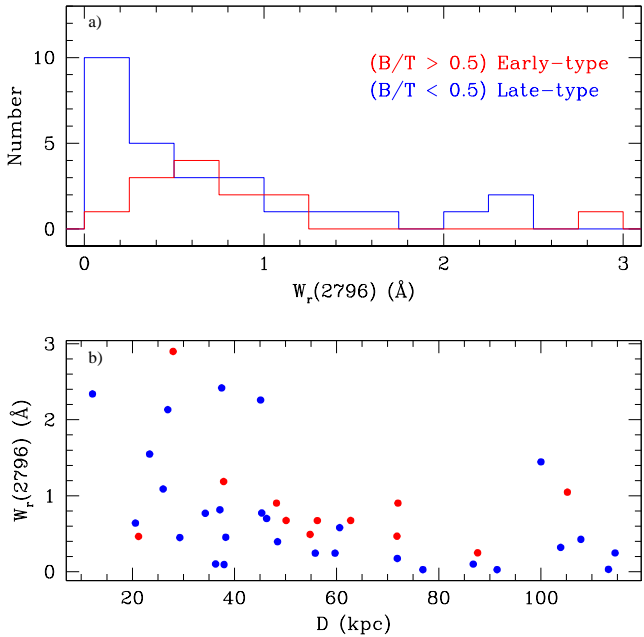}
\caption{--- (a) The distribution of the {\MgII} $\lambda 2796$
rest-frame equivalent width,$W_r(2796)$, as a function of galaxy
morphology. Galaxies have been separated into to classes using the
bulge-to-total ratio: late-type ($B/T>0.5$) and early-type
($B/T<0.5$). Note that a large fraction of late-type galaxies are
associated with low equivalent width systems.  --- (b) The impact
parameter as a function of the $W_r(2796)$ is shown for all 40
galaxies in our sample. The blue points are late-type galaxies and the
red points are early-type galaxies. We find a $3.2\sigma$
anti-correlation. Note the large scatter in the correlation and the
lack for high $D$ galaxies with high $W_r(2796)$.}
\label{fig:D}
\end{figure}

For our sample, we find $27$ late-type galaxies and $13$ early-type
galaxies; these numbers are consistent with the distribution of galaxy
morphologies found in the field environment
\citep{vdb04,hammer05,hammer09}. In Figure~\ref{fig:D}a, we show the
distribution of galaxy type as a function of $W_r(2796)$. The mean
$W_r(2796)$ for the early-type galaxies is 1.0~{\AA} and is slightly
lower for late-type galaxies at 0.7~{\AA}; both types are
accompanied by a large spread in equivalent width.

In Figure~\ref{fig:D}b, we show the $W_r(2796)$ as a function of
impact parameter, where red points are early-type galaxies and blue
points are late-type galaxies.  The mean impact parameter for each
morphology type is the same: 57~kpc.  There are some interesting trends to
note.  For $D\lesssim 40$~kpc, we find a higher proportion of
late-type galaxies (12/15) that span a large equivalent width range.
Beyond 40~kpc, we find more of a mixture of early and late-types, with
late-type galaxies dominating (14/25).  In this impact parameter
regime, we also see that late-type galaxies tend to have weaker
absorption than early-types.  The data suggests that late-type
galaxies are the dominate galaxy type selected by {\MgII} absorption
at all impact parameters.  We note that \citet{chen10} also find more
spirals in close proximity to the quasar line of sight and find a
factor of two more early-type galaxies at higher impact parameters ($D
> 43$~kpc).  A K-S test yields that the equivalent width distributions
of the late-type and early-type galaxies in our sample differ by
$2.1\sigma$.  Given the low number statistics of our sample and the
larger scatter of $W_r(2796)$ associated with late-type galaxies, we
caution any against any definitive conclusions being drawn from
Figure~\ref{fig:D}.

It could be possible that the physical scale lengths, i.e., $r_d$,
$r_h$, $r_b$, etc., of a galaxy may be related to its halo gas
absorption properties.  Having quantified the morphological parameters
of the galaxies in our sample, we have applied generalized Kendall and
Spearman rank correlation tests, which accounts for measured limits in
the sample \citep{feigelson85}, between the galaxy scale lengths,
S{\'e}rsic index, $n$, and bulge-to-total ratio, $B/T$ and the
measured absorption properties.  We find no evidence for trends with
these morphological parameters (all significance levels are less than
$2\sigma$).  This may suggest that galaxy mass, size, and/or global
stellar distribution is/are not important factor(s) in dictating the
absorption properties of gaseous halos.

\subsection{Impact Parameter and Absorption}
\label{sec:impact}

For many studies an anti-correlation between $W_r(2796)$ and $D$ has
been noted and employed to understand the geometry, mean radial
density profile, and cross-sectional covering fraction of {\MgII} halo
gas
\citep[e.g.,][]{lanzetta90,steidel95,csv96,mo96,bouche06,chelouche08,chen08,chen10,steidel10}.
For our sample, we find a $3.2\sigma$ anti-correlation (see
Figure~\ref{fig:D}b) .  Though it is clear that projected
galactocentric distance is a predominant parameter governing
absorption strength, there is significant scatter in the relation,
showing that distance from the host galaxy may not be the only
physical parameter governing the distribution and observed quantity of
halo gas.  For example, at various levels of influence, star formation
rate, environment, or the orientation of the line of sight (galaxy
inclination and/or position angle relative to the quasar) could also
play a role.

As a caveat with regards to interpreting this anti-correlation, we
note that, given the classical methods used to identify host galaxies
(find absorption then search for host galaxy), our sample and previous
samples are relatively biased toward galaxies in close proximity to
the quasar, which may artificially lead to the observed
anti-correlation.  As unbiased surveys
\citep[e.g.,][]{tripp-china,barton09} are enlarged, we will be able to
obtain more robust data to examine the significance of the $D$ and
$W_r(2796)$ anti-correlation.  Taken at face value, the observed
anti-correlation between $W_r(2796)$ and $D$ would suggest that either
the column density and/or velocity spread of the absorbing gas
decreases with galactocentric distance.

In Table~\ref{tab:KStest}, we present the results of additional
correlation tests between the absorption properties and $D$.  What is
immediately clear is that {\it all\/} our measures of the gas
absorption strength decrease with $D$, although we note that the
doublet ratio increases with $D$ as expected. $W_r(2803)$, which is
less sensitive to saturation than is $W_r(2796)$, anti-correlates
significantly with $D$.  Though $N_cl$, $N_vp$ and $\tau$ do not show
a significant anti-correlation with $D$, they do follow the trend of
decreasing with $D$.  Taken together, the anti-correlation trends show
a consistent signature that, on average, {\MgII} absorbing gas becomes
optically thinner with increasing projected galactocentric distance
from the host galaxy.
 
As stated above, the decreasing equivalent widths indicate that either
the column density and/or velocity spread of the absorbing gas
decreases with galactocentric distance.  However, the facts that (1)
$W_r(2803)$ shows a slightly higher significance than $W_r(2796)$, (2)
the doublet ratio increases with $D$, and (3) the mean velocity,
$W_{vs}$, and the velocity asymmetry, $A$, scatters with $D$ (shows no
signature of decreasing with $D$), together suggest that a decrease in
column density with $D$, rather than just a decrease in velocity
structure, is likely the dominant behavior of the halo gas.

\begin{table*}
\begin{center}
  \caption{Selected results of the Kendall and Spearman rank
correlation tests between the {\MgII} absorption properties and the
associated galaxy properties. In column (1) are the tested properties
and (2) has the number of systems tested. In columns (3)--(5) are the
Spearman correlation coefficient, $r_{_{\rm S}}$, the probability,
$P_{_{\rm S}}$, that the tested data are consistent with the null
hypothesis of no correlation, and the number of standard deviations,
$N_{_{{\rm S}\sigma}}$ where the ranks are drawn from a normal
distribution. In columns (6)--(8) are the Kendall $\tau_{_{\rm K}}$,
the probability, $P_{_{\rm K}}$, and $N_{_{{\rm K}\sigma}}$.}
  \vspace{-0.5em}
\label{tab:KStest}
{\footnotesize\begin{tabular}{lcrcrrccr}\hline
Tests & Galaxies & $r_{_{\rm S}}$ &$P_{_{\rm S}}$ & $N_{_{{\rm S}\sigma}}$ & $\tau_{_{\rm K}}$ & $P_{_{\rm K}}$ & $N_{_{{\rm K}\sigma}}$ \\\hline 
$  D$ vs. $    W_r(2803)  $   &    40 &    $-0.51$ &0.0015 &  3.18 &  $-0.73$  & 0.0010 & 3.30 \\
$  D$ vs.  $    W_r(2796) $   &    40 &    $-0.49$ &0.0022 &  3.06 &  $-0.70$  & 0.0014 & 3.19 \\
$  D$ vs.  $  DR          $   &    40 &    $ 0.37$ &0.0219 &  2.29 &  $ 0.51$  & 0.0210 & 2.31 \\
$  D$ vs.  $     N_{cl}    $  &    34 &    $-0.33$ &0.0578 &  1.90 &  $-0.46$  & 0.0503 & 1.96 \\
$  D$ vs.  $         N_{vp}$  &    34 &    $-0.32$ &0.0644 &  1.85 &  $-0.43$  & 0.0752 & 1.78 \\
$  D$ vs.  $       W_{vs}  $  &    34 &    $-0.28$ &0.1038 &  1.63 &  $-0.35$  & 0.1461 & 1.45 \\
$  D$ vs.  $        \tau   $  &    34 &    $-0.22$ &0.2082 &  1.26 &  $-0.29$  & 0.1492 & 1.44 \\
$  D$ vs.  $            N_a$  &    34 &    $-0.14$ &0.4202 &  0.81 &  $-0.21$  & 0.3076 & 1.02 \\
$  D$ vs.  $\left< V \right>$ &    34 &    $-0.08$ &0.6361 &  0.47 &  $-0.11$  & 0.6351 & 0.48 \\
$  D$ vs.  $         A     $ &    34 &    $ 0.01$ &0.9783 &  0.03 &  $-0.03$  & 0.9056 & 0.12 \\
              &	      &        	   &	   &   	    &  	     &	      &	      \\	     
$B-K$ vs. $     N_{cl}    $  &    23 &    $-0.47$ &0.0272 &  2.21 &  $-0.66$  & 0.0251 & 2.24 \\
$B-K$ vs. $         N_{vp}$  &    23 &    $-0.49$ &0.0207 &  2.31 &  $-0.65$  & 0.0301 & 2.17 \\
$B-K$ vs. $\left< V \right>$ &    23 &    $ 0.41$ &0.0554 &  1.92 &  $ 0.63$  & 0.0344 & 2.12 \\
$B-K$ vs. $       W_{vs}  $  &    23 &    $-0.33$ &0.1218 &  1.55 &  $-0.52$  & 0.0809 & 1.75 \\
$B-K$ vs. $    W_r(2796)  $  &    27 &    $-0.31$ &0.1156 &  1.57 &  $-0.45$  & 0.0991 & 1.65 \\
$B-K$ vs. $         A     $  &    23 &    $ 0.36$ &0.0963 &  1.66 &  $ 0.47$  & 0.1125 & 1.59 \\
$B-K$ vs. $        \tau   $  &    23 &    $-0.33$ &0.1248 &  1.54 &  $-0.37$  & 0.1489 & 1.44 \\
$B-K$ vs. $    W_r(2803)  $  &    27 &    $-0.29$ &0.1423 &  1.47 &  $-0.37$  & 0.1749 & 1.36 \\
$B-K$ vs. $  DR           $  &    27 &    $ 0.24$ &0.2139 &  1.24 &  $ 0.35$  & 0.2030 & 1.27 \\
$B-K$ vs. $            N_a$  &    23 &    $-0.21$ &0.3214 &  0.99 &  $-0.23$  & 0.3731 & 0.89 \\
                 &	     &        	   &	   &  	       &     	     &	      &   \\	     
$M_B$ vs. $            N_a$  &     34&    $ 0.20$ &0.2523 &  1.15 &  $ 0.26$  & 0.2292 & 1.20 \\
$M_B$ vs. $         A     $  &     34&    $-0.19$ &0.2791 &  1.08 &  $-0.24$  & 0.3277 & 0.98 \\ 
$M_B$ vs. $  DR            $ &     40&    $ 0.13$ &0.4211 &  0.81 &  $ 0.20$  & 0.3633 & 0.91 \\ 
$M_B$ vs. $       W_{vs}  $  &     34&    $-0.15$ &0.3795 &  0.88 &  $-0.21$  & 0.3736 & 0.89 \\
$M_B$ vs. $     N_{cl}    $  &    34 &    $-0.13$ &0.4639 &  0.73 &  $-0.18$  & 0.4515 & 0.75 \\
$M_B$ vs. $        \tau   $  &     34&    $ 0.12$ &0.4774 &  0.71 &  $ 0.16$  & 0.4625 & 0.74 \\
$M_B$ vs. $\left< V \right>$ &     34&    $ 0.08$ &0.6487 &  0.46 &  $ 0.11$  & 0.6564 & 0.45 \\
$M_B$ vs. $    W_r(2796)  $  &     40&    $-0.08$ &0.6222 &  0.49 &  $-0.09$  & 0.6919 & 0.40 \\
$M_B$ vs. $         N_{vp}$  &     34&    $ 0.05$ &0.7700 &  0.29 &  $ 0.09$  & 0.6998 & 0.39 \\       
$M_B$ vs. $    W_r(2803)  $  &     40&    $-0.08$ &0.6246 &  0.49 &  $-0.07$  & 0.7354 & 0.34 \\ 
              &	      &       	   &	   &  	       &     	     &	      &	      \\	     

$ i  $ vs. $     N_{cl}    $ &    34 &    $  0.38 $ &0.0297 &2.17  &  $  0.52$  &0.0279  &2.20\\
$ i  $ vs. $            N_a$ &    34 &    $  0.36 $ &0.0389 &2.07  &  $  0.48$  &0.0292  &2.18\\
$ i  $ vs. $         N_{vp}$ &    34 &    $  0.34 $ &0.0546 &1.92  &  $  0.51$  &0.0327  &2.14\\
$ i  $ vs. $        \tau   $ &    34 &    $  0.31 $ &0.0790 &1.76  &  $  0.40$  &0.0729  &1.79\\
$ i  $ vs. $    W_r(2796)  $ &    40 &    $  0.24 $ &0.1318 &1.51  &  $  0.33$  &0.1358  &1.49\\
$ i  $ vs. $         A     $ &    34 &    $ -0.31 $ &0.0799 &1.75  &  $ -0.36$  &0.1381  &1.48\\
$ i  $ vs. $    W_r(2803)  $ &    40 &    $  0.19 $ &0.2251 &1.21  &  $  0.28$  &0.2039  &1.27\\
$ i  $ vs. $       W_{vs}  $ &    34 &    $  0.20 $ &0.2412 &1.17  &  $  0.28$  &0.2474  &1.16\\
$ i  $ vs. $\left< V \right>$&    34 &    $ -0.05 $ &0.7794 &0.28  &  $ -0.07$  &0.7668  &0.30\\
$ i  $ vs. $  DR            $&    40 &    $ -0.04 $ &0.8161 &0.23  &  $ -0.04$  &0.8521  &0.17\\
                      &	      &        	   &	   &  	    &       	     &	      &	      \\	     
$ i/D$ vs. $    W_r(2796)  $ & 39    &    $  0.63$ &  9.24$\times 10^{-5}$ &  3.91  &  $  0.97$  &1.40$\times 10^{-5}$ &4.34 \\
$ i/D$ vs. $    W_r(2803)  $ & 39    &    $  0.63$ & 0.0001 &  3.88  &  $  0.96$  &1.84$\times 10^{-5}$ &4.28 \\
$ i/D$ vs. $            N_a$ & 33    &    $  0.65$ & 0.0002 &  3.70  &  $  0.93$  &3.79$\times 10^{-5}$ &4.12 \\
$ i/D$ vs. $        \tau   $ & 33    &    $  0.61$ & 0.0006 &  3.44  &  $  0.87$  &0.0001 & 3.86 \\
$ i/D$ vs. $     N_{cl}    $ & 33    &    $  0.58$ & 0.0010 &  3.30  &  $  0.85$  &0.0004 & 3.54 \\
$ i/D$ vs. $         N_{vp}$ & 33    &    $  0.53$ & 0.0029 &  2.98  &  $  0.78$  &0.0015 & 3.18 \\
$ i/D$ vs. $         DR    $ & 39    &    $ -0.41$ & 0.0115 &  2.53  &  $ -0.58$  &0.0093 & 2.60 \\
$ i/D$ vs. $       W_{vs}  $ & 33    &    $  0.40$ & 0.0234 &  2.27  &  $  0.56$  &0.0227 & 2.28 \\
$ i/D$ vs. $         A     $ & 33    &    $ -0.08$ & 0.6357 &  0.47  &  $ -0.13$  &0.6090 & 0.51 \\
$ i/D$ vs. $\left< V \right>$& 33    &    $  0.03$ & 0.8552 &  0.18  &  $ -0.01$  &0.9629 & 0.05 \\\hline
\end{tabular}}
\end{center}
\end{table*}


\subsection{Galaxy Color and Absorption}

Given the results of \citet{zibetti07}, who statistically show that
larger $W_r(2796)$ is associated with bluer galaxies, and the results
of \citet{menard09}, who find a highly significant correlation between
{\OII} luminosity and $W_r(2796)$, we might expect to find that bluer
galaxies in our sample are associated with larger $W_r(2796)$.  We
thus tested for a correlations between $B-K$ galaxy color and
absorption properties in our sample.

At the range of significance level $\sim$2--2.2$\sigma$, dominated by
$N_{cl}$, $N_{vp}$, and $\left<V \right>$, we find larger absorption
quantities tend to be associated with the bluer galaxies.  The
Spearman Kendall results are presented in Table~\ref{tab:KStest}.  If
we were to interpret these trends, we would infer that redder
(early-type) galaxies are associated with low column density
absorption with fewer kinematic components than bluer (late-type)
galaxies.  We do not find evidence supporting a trend of increasing
$W_r(2796)$ with decreasing $B-K$, as found by \citet{zibetti07}.

The result for our sample is consistent with the results of
\citet{chen10}, who examined a sample of 71 systems.  They find no
significant correlation between $B-R$ galaxy color and $W_r(2796)$.
It remains difficult to directly compare our work and the work of
\citeauthor{chen10} because the works employ different colors.  It is
also difficult to compare our work to those of \citeauthor{zibetti07}
due to the stacking nature of their data analysis, the difference in
the $W_r(2796)$ distribution of the two samples, and the use of
different colors in both samples.  We further discuss these samples
and their results in Section~\ref{sec:dis}.

\subsection{Galaxy Luminosity and Absorption}

A Holmberg-like luminosity scaling of the radius of {\MgII} absorbing
halo sizes has been discussed at length in the literature
\citep{steidel95,chen08,kacprzak08,chen10}.  For the sample of
galaxies presented here, \citet{kacprzak08} showed that the {\MgII}
absorption halo size exhibits a scaling with the galaxy luminosity
(albiet not with a clean cut-off size).  Do the absorption properties
also exhibit a scaling with luminosity?  To investigate this question,
we examined if there is a correlation between absorption properties
and $M_B$ (effectively the galaxy luminosity).  We find no evidence of
a correlation with $M_B$ for any the absorption properties.  The
Spearman and Kendall results are presented in Table~\ref{tab:KStest}.
This would suggest that galaxy luminosity is not a predominant
governing factor in determining the absorption strength, optical
thickness, or velocity spread of the {\MgII} gas in our sample.

\subsection{Galaxy Orientation and Absorption}
\label{sec:orientation}

\begin{figure*}
\includegraphics[angle=0,scale=0.80]{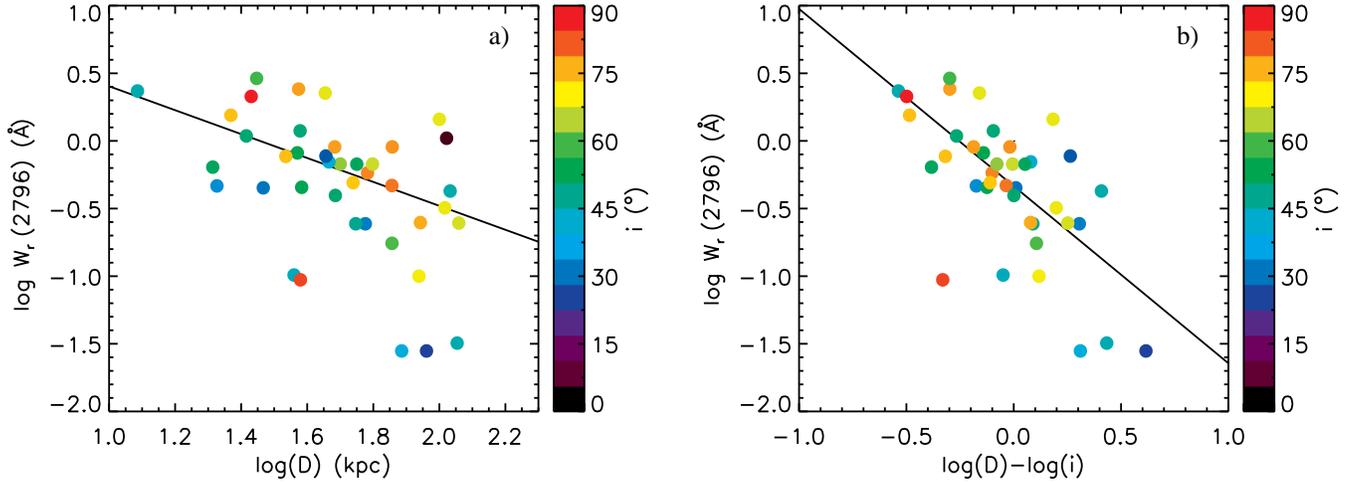}
\caption{--- (a) The 3.2$\sigma$ anti-correlation between $W_r(2796)$
and impact parameter $D$. Note the large scatter in the
correlation. The dark line shows a maximum-likelihood fit to the data
with log$(W_r(2796))= 1.228-0.884$log$(D)$. The points are color-coded
as a function of galaxy inclination angle, $i$.  It is apparent that
galaxies with high, mid, and low inclinations populate different
regions of the plot. For a fixed $D$, high inclination galaxies tend
to have higher $W_r(2796)$ than low inclination galaxies. This
inclination gradient in the $W_r(2796)$ direction suggests that
inclination plays a secondary but significant role in determining the
$W_r(2796)$ of an absorption system. --- (b) A correction for
inclination is applied to the $W_r(2796)$ and $D$
anti-correlation. The scatter is now reduced and results in a
4.3$\sigma$ correlation between $W_r(2796)$ and $i/D$. Note the
different inclinations are now overlapping compared to (a). The dark
line shows a maximum-likelihood fit to the data with log$(W_r(2796))=
-0.334-1.310$log$(D/i)$. }
\label{fig:D2}
\end{figure*}


\citet{lanzetta92} and \citet{cc96} showed that if there are preferred
systematic kinematics and spatial distributions of {\MgII} absorbing
gas relative to the host galaxies, then the absorption strengths and
kinematics would be expected to follow a predictable behavior as a
function of galaxy orientation and impact parameter.  By orientation,
we refer to the galaxy inclination, $i$, and the position of the
quasar line of sight relative to the major axis of the galaxy,
$\theta$, which ranges from $\theta = 90^{\circ}$ when the quasar
aligns with the major galaxy axis to $\theta = 0^{\circ}$ when the
quasar aligns with the galaxy minor axis.

\citet{cc96} further demonstrated that {\MgII} halos could be
disk-like since, statistically, the {\MgII} redshift path density could
be satisfactorily explained by extended disks as well as by spherical
halos.  They predicted that if absorbers are disk-like in origin, then
(1) the mean $W_r(2796)$ for a given inclination increases with $i$,
(2) the relative number of stronger absorbers peaks when the quasar
probes the major axis ($\theta = 0^{\circ}$), and (3) the mean ratio
of the absorption kinematic spread to the Tully Fisher velocity of the
galaxy increases with $i$. Making a direct comparison of galaxy and
{\MgII} absorption kinematics, \citet{steidel02} showed that galaxy
halos are partially understood by lagging disk-like kinematics.
Binning {\MgII} absorption profiles by inclination,
\citet{kacprzak10a} found that the mean optical depth and mean
velocity spread of the absorbing gas was larger for higher inclination
galaxies.

From the above predictions, if flattened halos that are co-planer with
the galaxy disk dominate, we expect the absorption properties to
correlate with inclination and anti-correlate with position angle (be
strongest at $i=90^{\circ}$ and $\theta=0^{\circ}$).  On the other
hand, it is also possible that star-formation driven winds could be a
dominant source of observed {\MgII} absorption \citep[especially for
the stronger absorbers, e.g.,][]{zibetti07,weiner09}.  In contrast to
AGN driven winds, star-formation driven winds are spatially
distributed geometrically perpendicular to the plane of the galaxy
\citep[see][]{veilleux05}.  Thus, for winds, we would expect the
absorption properties and velocity widths to be anti-correlated with
inclination and correlate with position angle (be strongest at
$i=0^{\circ}$ and $\theta=90^{\circ}$).  In reality, both scenarios,
possibly further complicated by additional scenarios such as mergers,
IGM filament accretion, etc., contribute to the presence of {\MgII}
absorption.  Are the predicted observational signatures of one of
these scenarios manifest in the data, which would suggest predominance
of the scenario?

GIM2D yields two position angles -- one for the bulge and one for the
disk.  We employ the bulge-to-total ratio for adopting the best
representative galaxy position angle, such that when $B/T>0.5$ we use
bulge, $\theta_b$, and when $B/T<0.5$ we use the disk, $\theta_d$. For
our sample, we find that the distribution of inclinations and position
angles of galaxies selected by {\MgII} absorption are not inconsistent
with having been drawn from a population of field galaxies (for which
the distribution of inclinations is $\propto \sin^2i$ and position
angles are random).  A K-S test for the inclination distribution
yields a K-S statistic of 0.16 and ${\rm P(KS)} = 0.27$ and for the
position angle distribution yields 0.16 with ${\rm P(KS)} = 0.28$.  At
face value, this would suggest that galaxies chosen by known
absorption are distributed on the sky no differently than galaxies
selected at random.  This might further suggest that halos have a
uniform spatial distribution around their host galaxies.  However,
there is a wide range of absorption properties (which are governed by
the column densities and kinematics, i.e., velocity widths) and the
overall distribution of orientations relative to the line of sight
cannot, by themselves, discriminate whether absorption properties vary
in a systematic fashion with orientation.

 We find no correlation between $\theta$ and any of absorption
properties (the highest significance is 1.4$\sigma$); the absorption
properties are consistent with a random distribution as a function of
position angle (these results are independent of our $B/T$ selection
criteria).

The results of correlation tests between the absorption properties and
inclination $i$ are similar to the tests with impact parameter; we
find several trends toward a correlation above the $2.0\sigma$ level.
The Spearman and Kendall quantities are presented in
Table~\ref{tab:KStest}.  The strongest trends are $N_{cl}$, $N_a$, and
$N_{vp}$ all showing that these quantities tend to increase with
increasing inclination.  Taken together, these trends could suggest a
{\it potential\/} correlation between disk inclination and the column
density, number of clouds, and $W_r(2796)$.

Overall, these results are suggestive (not definitive) that the
observational signature from the predictions of a co-planer geometric
model dominate over the other plausible scenarios for the origin and
location of the absorbing gas.  They are consistent with the findings
of \citet{kacprzak10a} that the optical depth and velocity spread of
the gas, relative to the galaxy systemic velocity, increases with disk
inclination.  Interestingly, we do not find a trend suggested by
\citet{cc96} for increasing $W_{vs}$ (normalized by the Tully Fisher
velocity, computed from $L_B$) with increasing inclination
(0.2$\sigma$).  A simple galaxy disk model is not supported by the
position angle data, which are not necessarily consistent with the
co-planer geometry scenario.  If the absorbing gas were in a disk,
then the radial distance out on the disk mid-plane probed by the line
of sight is given by $r = D\sqrt{1+\tan^2i \sin^2 \theta}$.
Correlation tests between the absorption properties and $r$ might be
expected to have higher significance (less scatter) than those with
$D$ (even allowing for all disks not being equal).  For our sample,
the absorption properties exhibit uncorrelated scatter with $r$ (no
test had a significance above $2\sigma$).  Thus, at best, we find the
data suggestive that most {\MgII} absorbers have a co-planer geometry,
but are not necessarily disks \citep[also see the geometric and
kinematic models of][]{kacprzak10a}. Since there is a substantial
scatter in the absorption properties, there is some difficulty in
interpreting these trends.  Since the distribution of inclinations for
our sample is consistent with that of a random sample of galaxies, our
sample has a larger number of higher inclination galaxies than lower
inclination galaxies [18/40 galaxies have $i > 60^{\circ}$ and only
5/22 have $i < 30^{\circ}$).  With a paucity of low inclination
(predominantly face on) galaxies, any possible scatter in the
absorption properties for low inclination may not be well represented
in our sample.  However, there is no clear or evident bias in our
sample that could explain why the lower inclination galaxies have
lower strength absorption properties.

\subsection{Normalization by Impact Parameter}
\label{sec:normalize}

As discussed in Section~\ref{sec:impact}, the anti-correlations
between $W_r(2796)$ and $D$ and between $W_r(2803)$ and $D$, and
decreasing trends with the remaining absorption properties, all
corroborate a physical picture in which the column densities diminish
with increasing $D$.  This suggests that impact parameter has a strong
influence on the absorption properties and that calibrating out impact
parameter for the correlation tests between galaxy properties and
absorption properties may uncover otherwise diluted physically
motivated trends or correlations.  From the point of view of
non-parametric rank correlation estimators, multiplying the absorption
properties by $D$ is equivalent to dividing the galaxy properties by
$D$.  We adopt the latter approach.  Since one of the strongest
indicators of a dependence of the absorption on the galaxy properties
is orientation, we focused on the galaxy inclination.\footnote{We
performed the correlation tests on all galaxy properties, but find
that, unless discussed in the text, there were no other statistically
significant (i.e., greater than $3\sigma$) correlations when
normalizing by $D$.}$^,$\footnote{We have removed galaxy Q2206-199G2
from the rank correlation tests when normalizing by $D$ because its
value of inclination and errors, $i=2.9^{+17.4}_{-2.9}$ and
$D=105.2$kpc, provides no constraints on the ratio of log$(D/i)$.}

Following normalization of $i$ by $D$, we find that $N_{vp}$,
$N_{cl}$, $\tau$, $N_a$, $W_r(2803)$, and $W_r(2796)$ are all
correlated with the normalized inclination, $i/D$, at a greater than
$3.2\sigma$ significance.  The strongest correlation is between
$W_r(2796)$ and $i/D$, which has a one part in $10^5$ probability of
being consistent with no correlation ($4.3\sigma$).  The results are
presented in Table~\ref{tab:KStest}.  We note that $D$ does not appear
to correlate with $i$ (only at 1.5$\sigma$).

\begin{figure}
\includegraphics[angle=0,scale=0.80]{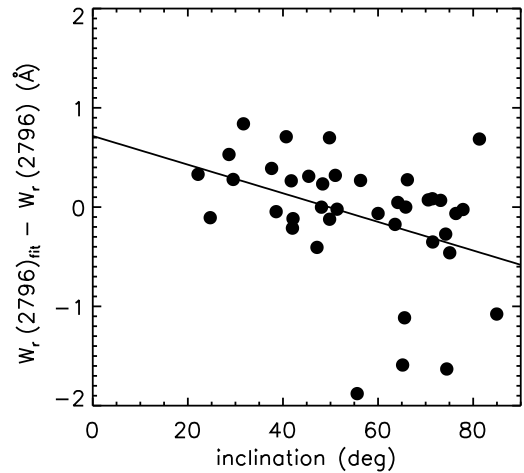}
\caption{The galaxy inclination as a function of the equivalent width
residuals computed from the fit between $W_r(2796)$ versus $D$
(Figure~\ref{fig:D2}a). We find a 2.6$\sigma$ correlation between the
residuals (scatter) and the galaxy inclination. This correlation
further demonstrates the significance in the reduction of scatter
between $W_r(2796)$ versus $D$ when galaxy inclination is taken into
account.}
\label{fig:prob} 
\end{figure}

In Figure~\ref{fig:D2}a, we show the $3.2\sigma$ anti-correlation
between $W_r(2796)$ and $D$. As we have discussed in
Section~\ref{sec:impact}, there is large scatter observed in the
$W_r(2796)$ and $D$ anti-correlation. In Figure~\ref{fig:D2}a, we also
color-code the data points as a function of galaxy inclination.  It is
clearly noticeable that high, mid, and low galaxy inclinations
populate different regions of the plot and are offset from each other
as a function of $W_r(2796)$ (i.e., for a fixed $D$, high inclination
galaxies tend to have higher $W_r(2796)$ than low inclination
galaxies). The galaxy inclination gradient in the $W_r(2796)$
direction suggests that inclination plays a secondary role but
significant determining the $W_r(2796)$ of an absorption system.

In Figure~\ref{fig:D2}b, we apply a galaxy inclination correction. We
note that our discussions of rank correlation tests are in terms of
the galaxy inclination normalized by the impact parameter, $i/D$,
however, in Figure~\ref{fig:D2}b we chose to plot $D/i$ in order to
aid the reader to directly compare the scatter from both panels. Note
that the scatter is significantly reduced and that there is now no
clear inclination gradient seen in the distribution of data.  A
maximum-likelihood fit to the $W_r(2796)$ and $i/D$ correlation is
provided in Table~\ref{tab:allcorr}. In Figure~\ref{fig:prob} we show
that if we fit the anti-correlation between $W_r(2796)$ versus $D$
(Figure~\ref{fig:D2}a), then the equivalent width residuals computed
from the fit correlate with galaxy inclination at $2.6\sigma$, further
supporting the significance in the reduction of scatter in
Figure~\ref{fig:D2}b.

We also apply bootstrap statistics to test the probability that a
random distribution of inclination values could produce the
inclination gradient as a function of $W_r(2796)$ seen in
Figure~\ref{fig:D2}a, which results in the tightening of the
anti-correlation seen in Figure~\ref{fig:D2}b.  We used our 40 data
points of $W_r(2796)$ and $D$ pairs since they exhibit the expected
correlation shown in Figure~\ref{fig:D2}a. We then took our measured
values of $i$ and randomly reassigned them to the $W_r(2796)$ and $D$
pairs, thereby creating a new bootstrap realization of the sample. We
then repeated this process $1\times10^6$ times and computed the
Kendall rank correlation test for each realization of the sample.  We
find that the probability of this correlation occurring by chance due
to random selection of galaxy inclinations is $P=0.00040$. Thus, it is
unlikely that this correlation is due to random chance at the
$3.54\sigma$ significance level.

Thus, we find that the distribution of halo gas absorption strengths
has both an impact parameter \emph{and} inclination dependence, which
suggests that the gas in the halo is not spherically distributed.
That is, these results suggest that {\MgII} halo gas has a co-planer
geometry and is coupled to the inclination of the galaxy disk. The
correlation is opposite to that expected for wind models.

We also examined if multiples of the galaxy bulge and disk scale
lengths, $D/r_b$ and $D/r_d$, and half-light radii, $r_h$, might
further reduce the scatter in the above correlations by using the
quantities $i/(D/r_d)$, etc.  However, we found that multiples of the
scale lengths resulted in greater scatter and no statistically
significant correlations with absorption properties. We also
normalized the galaxy position angle by $D$ and re-examine the scatter
in the absorption properties with $\theta$.  We found a slight
increase in the significance levels from $1.4\sigma$ and below to
$\simeq 3.0\sigma$ and below.  However, these significance levels
remain lower than the significance level of the anti-correlation
between $W_r(2796)$ and $D$.  As such, the increased significance
levels are primarily driven by the normalization of $D$.  The upshot
is that, if the selection of galaxies by {\MgII} absorption yields a
preferred position angle--impact parameter range in relation to the
absorption strength properties, it has a much larger scatter than does
the preferred inclination--impact parameter range for a given range of
absorption properties.

\section{Discussion}
\label{sec:dis}

\begin{table*}
\begin{center}
  \caption{Currently known quantified correlations between the {\MgII}
absorption properties and the host galaxy properties: (1) The
correlation between $W_r(2796)$ and galaxy inclination normalized by
impact parameter, $i/D$; (2) The anti-correlation between $D$ and
$M_B$ with $W_r(2796)$; (3) The correlation between {\OII} luminosity
surface density ($\Sigma$L$_{\OII}$) and $W_r(2796)$; (4) The
correlation between galaxy asymmetry ($A$), normalized by $D$, and
$W_r(2796)$; (5) The anti-correlation between the host galaxy halo
mass ($M_{halo}$) and $W_r(2796)$. \citet{bouche06} did not provided
parametrized relationship for their correlation, however we used the
binned data from their Table~3 and applied a least least squares
fit. }
  \vspace{-0.5em}
\label{tab:allcorr}
{\footnotesize\begin{tabular}{lccccll}\hline

Properties                        & Redshift &$W_r(2796)$ & (Anti-)             & Signif-       &  Parametrized Relationship                           & Reference     \\
                                  & Range    &Range (\AA) &  Correlation        & icance        &                                                        &      \\\hline 
$i/D$ \& $W_r(2796)$              & 0.3--1.0 &  0.03--2.9 &  Correlation        & $4.3\sigma$  &  log$(W_r(2796))= -0.334-1.310$log$(D/i)$                               &   This paper.     \\
                                  &          &            &                     &              &                                                           &                    \\[-2.0ex]
$D + M_B$ \& $W_r(2796)$          & 0.1--0.5 &  0.1--2.4  &  Anti-correlation   &  $7\sigma$   & log$(W_r(2796))=A$log$(D)-B(M_B-M_B^{\ast})+ C$            &  \citet{chen10}    \\                 
                                  &          &            &                     &              &       $A=-1.93\pm0.11$, $B=-0.27\pm0.02$                  &                    \\
                                  &          &            &                     &              &                      $C=2.51\pm0.16$                      &                    \\
                                  &          &            &                     &              &                                                           &                    \\[-2.0ex]
$\Sigma$L$_{\OII}$ \& $W_r(2796)$ & 0.4--1.3 &  0.7--6.0  &  Correlation        &  $15\sigma$  &  $\Sigma$L$_{\OII} =$$A$($W_r(2796)$/1\AA)$^\alpha$       &  \citet{menard09}  \\
                                  &          &            &                     &              & $A=1.48\pm0.18\times10^{37}$erg s$^{-1}$ kpc$^2$          &                    \\
                                  &          &            &                     &              & $\alpha=1.75\pm0.11$                                      &                    \\
                                  &          &            &                     &              &                                                           &                    \\[-2.0ex]
$A/D$ \& $W_r(2796)$       & 0.3--1.0 &  0.03--2.9&  Correlation        &  $3.3\sigma$ & $A/D = 2.48\times10^{-3}W_r(2796) + 8.15\times10^{-4}$   &  \citet{kacprzak07}\\
                                  &          &            &                     &              &                                                           &                    \\[-2.0ex]
$M_{halo}$ \& $W_r(2796)$         & 0.4--0.8 &  0.5--5.0  & Anti-correlation    &   $\cdots$   & log$(M_{halo})=13 - 0.6W_r(2796)$                         &  \citet{bouche06}  \\\hline
 
\end{tabular}}
\end{center}
\end{table*}

Several studies have shown that the morphologies of intermediate
redshift ($0.3<z<1.0$) {\MgII} absorption-selected galaxies appear to
be qualitatively similar to those of ``typical'' local field galaxies
\citep{steidel98,chen01,chen03,kacprzak07}.  Using GIM2D to quantify
the morphological parameters of $0.3 < z < 1.0$ {\MgII}
absorption-selected galaxies, we find that their bulge and disk size
distributions are also similar to those found for field galaxies at
similar redshift.  Our results quantitatively show that galaxies
selected by {\MgII} absorption are, broadly speaking, ``typical''
galaxies.  Nevertheless \citet{kacprzak07} has shown that {\MgII}
absorption-selected galaxies appear to have higher morphological
asymmetries than field galaxies.  If the latter is true, then
accurately modeling absorption-selected galaxies with smooth light
profiles may be more difficult given their higher level of
morphological asymmetries. The difficulties may induce more scatter in
the distribution of morphological parameters thereby potentially
mitigating the significance levels of any correlations between galaxy
properties and absorption properties.

It remains strongly debated whether {\MgII} absorption systems arise
from star formation driven winds or an array of structures such as
tidal streams, satellites, filaments, etc. Here we argue that our
sample of galaxies, 75\% of which have $W_r(2796)< 1.0$~{\AA}, does
not support a scenario in which winds are the predominant mechanism
producing the absorption.  Overall, it is likely that an admixture of
these processes contribute to the observed {\MgII} absorption systems;
below, we will argue that winds may dominate the high equivalent width
regime, whereas other processes dominate the lower equivalent width
regime. 

For reference, in Table~\ref{tab:allcorr} we list the known quantified
correlations between the {\MgII} absorption properties and the host
galaxy properties, which we will be discussing in the remainder of
this section. The $W_r(2796)$ range used in each study is also listed.

Studies of strong {\MgII} systems (i.e., $W_r(2796) \gtrsim 1$~{\AA}),
where host galaxies typically have $D\lesssim 25$~kpc, have shown that
they are likely produced by winds from high star-forming galaxies
\citep[e.g.,][]{prochter06}.  \citet{bouche06} report an
anti-correlation between $W_r(2796)$ and the amplitude of the
cross-correlation between luminous red galaxies (LRGs) and {\MgII}
absorbers \citep[also see][]{gauthier09,lundgren09}. They conclude
that stronger absorption systems are not produced by virialized gas
within galaxy halos but that they originate in supernovae-driven
winds. The \citeauthor{bouche06} results are consistent with those of
\citet{zibetti07}, who find that stronger absorption systems with
$W_r(2796) \gtrsim 1.2$~{\AA} are associated with blue star-forming
galaxies.  \citet{menard09} showed that for stronger systems,
$W_r(2796)$ correlates with the associated {\OII} luminosity, an
estimator of star formation rates \citep[also see][]{noterdaeme10}. In
a detailed study of two ultra-strong {\MgII} absorption systems, with
$W_r(2796) > 3.6$~{\AA}, \citet{nestor10} reported nearby galaxies
with star formation rates exceeding
$\sim10$--$90$~M$_{\odot}$~yr$^{-1}$ \citep[also
see][]{nestor07}. These results support the idea that winds produced
by star-forming regions and/or supernovae are responsible for a large
proportion of the strong absorption systems ($\gtrsim 1$~{\AA})
detected in the halos of galaxies.

Our study, and others
\citep[e.g.,][]{csv96,cv01,steidel02,cvc03,chen08,chen10,chen10b,kacprzak10a},
use higher resolution spectra that are sensitive to lower equivalents
widths. The general conclusions from these moderate-to-low equivalent
width studies all point to other sources for producing the observed
{\MgII} absorption.

Using a similar sample to the one presented here, \citet{kacprzak07}
found a correlation between galaxy asymmetry, normalized by $D$, and
$W_r(2796)$ suggesting that galaxy minor mergers and harassments may
be producing the absorption detected in halos. Their correlation
strengthens when $W_r(2796)\geq 1.4$~{\AA} systems were removed,
suggesting that interactions (or processes that gently perturb galaxy
morphology) may dominate the {\MgII} absorption profiles in the lower
equivalent width regime.  For our sample, we find trends that weaker
absorption properties are associated with redder galaxies; but there
is no suggested trend directly with $W_r(2796)$ as found for higher
$W_r(2796)$ systems as reported by \citet{zibetti07}.  The lack of
statistical significance may be due to the small number of galaxies in
our sample.  For a sample of 71 absorption selected galaxies with a
$W_r(2796)$ distribution similar to our sample, \citet{chen10} also
did not find a significant correlation between $W_r(2796)$ and galaxy
color.  The discrepancy between \citeauthor{zibetti07} and our work
and \citeauthor{chen10} may indicate that there is a fundamental
difference in the galaxies selected by stronger and weaker {\MgII}
absorption systems; weaker systems are possibly not correlated with
host galaxy star-formation rates or the luminosity-weighted stacking
procedure is not fully understood. However, this correlation may have
a substantial scatter, so that large samples, on the order of those
used by \citeauthor{zibetti07}, are required to obtain statistical
significance.


The {\MgII} equivalent widths, and the other measures of the {\MgII}
column density, decrease with $D$, indicating that halos exhibit a
natural decrease in gas column density with increasing projected
galactocentric distance.  This anti-correlation alone cannot be
leveraged to differentiate between wind scenarios and other {\MgII}
gas producing mechanisms; however, we would expect a difference in the
halo gas geometry for the competing scenarios (see
\S~\ref{sec:orientation} and \S~\ref{sec:normalize}).  Our sample
exhibits trends for increasing absorption properties (and scatter)
with higher inclination galaxies.  Accounting for the decreasing gas
column density (and its scatter) with increasing $D$ reduces the
scatter and yields a strong correlation ($4.3\sigma$).  At the very
least, this suggests that the combined effect of inclination and
impact parameter is such that weaker absorption is found at larger $D$
and smaller $i$ (far from face-on galaxies).  As $i$ increases for a
given $D$, we see $W_r(2796)$ increases, or as $D$ decreases for a
given $i$, we see $W_r(2796)$ increases.  Given that the velocity
widths, $W_{vs}$, do not follow this behavior, the data favor an
increasing column density (path length) for increasing inclination at
a fixed $D$.  Such behavior is expected for co-planer geometry with
kinematics that are not strictly coupled to disk rotation, but is not
what is expected for a wind geometry.  For a simple wind scenario, an
anti-correlation between $W_r(2796)$ and $i/D$ would be expected.
These trends are not evident in our data.

Recent cosmological SPH simulations of \citet{stewart11} have shown
that gas-rich mergers and cold-flow streams can produce a
circum-galactic cool gas component that predominately infalls towards
the host galaxy disk. They note that this gas accretion should be
observed as a relative host-galaxy/halo-gas velocity offsets of $\sim
100$~{\kms}. These offsets have already been directly observed at
$z\sim 0.1 - 1.0$ using {\MgII} absorption systems \citep{steidel02,
kacprzak10a, kacprzak11}. In such a scenario one would expect a
correlation between host galaxy inclination and absorption strength,
column density, etc., if the circum-galactic component was a pure
extension of the host galaxy disk. According to these simulations,
\citet{stewart11} found that in most cases the accreting gas
co-rotates with the central disk in the form of a warped extended cold
flow disk, and the observed velocity offset are in the same direction
as galaxy rotation. These models support the correlations between the
absorption properties and $i/D$ found here, and the warps observed in
the simulations may further explain the scatter seen in these
correlations.


The relative gas--galaxy kinematics also differ between low and high
equivalent width systems.  \citet{bond01b} studied the kinematics
$W_r(2796) >1.8$~{\AA} systems and found characteristics of wind
driven gas (double quasi-symmetric absorption over $\sim 400$~{\kms},
not characteristic of {\MgII} in DLAs).  On the other hand, systems
with $W_r(2796) \leq 1$~{\AA} are characterized by velocity widths no
greater than $\sim 50$~{\kms} accompanied by one to a few very weak
and narrow components and are not suggestive of wind kinematics
\citep{cv01,cvc03}.  For predominantly $W_r(2796) \leq 1$~{\AA}
systems, \citet{chen10} found that the velocity differences between
the {\MgII} absorption and the host galaxies are roughly 16~{\kms}
with a dispersion of 137~{\kms}.  These velocity offsets are much
lower than expected for wind driven outflows.

In fact, a direct kinematic comparison of six galaxies and their
{\MgII} absorption kinematics, with velocity offsets similar to those
of \citeauthor{chen10}, were shown to be fairly well described as
having lagging disk-like kinematics, though not all of the velocity
spread could be successfully modeled \citep{steidel02}.  However, the
discrepancies were on the order of $30$~{\kms} in the most extreme
cases. For a similar sample, \citet{kacprzak10a} also showed that,
even if the {\MgII} absorption tends to reside fully to one side of
the galaxy systemic velocity and aligns with one arm of the rotation
curve, not all the absorbing gas kinematics can be explained by a
co-rotating halo model.  Using cosmological simulations, they further
showed that even if the majority of the simulated {\MgII} absorption
arises is an array of structures, such as filaments and tidal streams,
the halo gas often has velocities consistent with the galaxy rotation
velocity. The line of sight gas motions along these structures in the
simulations reproduce a velocity distribution consistent with that
reported by \citet{chen10}.

Finally, it is also interesting to note the equivalent width redshift
path density evolution of {\MgII} absorbers: higher equivalent width
systems evolve in that there are fewer high $W_r(2796)$ systems per
unit redshift at low redshifts than at higher redshifts.  The
evolution is more pronounced for the highest $W_r(2796$) systems,
i.e., those with 2~{\AA} and above. As lower $W_r(2796)$ systems are
examined, the evolution weakens such that for all systems with
$W_r(2796) >0.3$~{\AA} the absorber population is consistent with the
no-evolution expectation for the presently accepted cosmology
\citep{steidel92,nestor05}.  Assuming winds are a transient phenomenon
in most galaxies, and given that the global star formation rate of the
universe decreases toward low redshifts \citep{madau98}, it would seem
reasonable that, globally, the higher equivalent width systems have
some causal connection to star formation rates.  On the other hand,
the fact that lower equivalent width systems do not evolve with
redshift would indicate that the processes producing lower
$W_r(2796)$, though varied, are fairly constant with cosmic time.
Though major merger rates are expected to decrease strongly, $\propto
(1+z)^{2.1}$, with redshift \citep[cf.,][]{stewert09}, it is not clear
that other scenarios such as minor mergers and/or accretion of the IGM
(filaments, etc.)  evolve as rapidly.  It could be that the lack of
redshift evolution of lower $W_r(2796)$ absorbers is linked to the
latter, possibly more ubiquitous processes.

The different physical characteristics and evolution of the
populations of high and low $W_r(2796)$ {\MgII} systems suggest
different physical mechanisms giving rise to each population.  The
correlation between $W_r(2796)$ and $i/D$ for our lower $W_r(2796)$
sample indicates that winds do not dominate the lower equivalent width
regime.  It appears that galaxy inclination plays a strong role in
determining the optical depth of the halo gas once the decreasing
column density with increasing impact parameters is taken into
account.  We find that edge-on systems are likely to produce higher
optical depth absorption systems, and lower optical depth absorption
systems are produced by face-on galaxies: this is the opposite effect
expected for a wind scenario. These results support a picture where
the {\MgII} absorption arises in structures that are relatively
co-planer to the host galaxy disk.  Such structures might include
accreting filaments or tidal streams from minor mergers in the galaxy
plane, and disk warps. These results are consistent with recent
cosmological simulations of \citet{stewart11} who find that the
accretion of cool gas via filaments and gas rich merges does in fact
form a stable disk that supplies gas and angular momentum to the host
galaxy.

\section{Conclusions}
\label{sec:conclusion}



We have performed a detailed study of a sample of 40 {\MgII}
absorption-selected galaxies between $0.3<z<1.0$.  The galaxies have
B-band absolute magnitudes that range between $-18.6 \leq M_B \leq
-23.8$ and are associated with {\MgII} absorption systems with
rest-frame equivalent widths that range between $0.03\leq W_r(2796)
\leq 2.9$~\AA. The {\MgII} absorption profiles were obtained from
HIRES/Keck and UVES/VLT quasar spectra; we did not have high
resolution spectra for six systems and use equivalent width
measurements from the literature.  The galaxies are at projected
separations of $12 \leq D \leq 115$~kpc from the quasar line-of-sight.

We have used GIM2D to model WFPC-2/{\it HST} images and extract
quantified morphological parameters for 40 {\MgII} absorption-selected
galaxies. These parameters include: the bulge-to-total fraction
($B/T$), the bulge semi-major axis effective radius ($r_b$), the bulge
ellipticity ($e_b$), the bulge position angle ($\theta_b$), the bulge
S{\'e}rsic index ($n$), the semi-major axis disk scale length ($r_d$),
the disk inclination ($i$), and the disk position angle
($\theta_d$). These properties help us further compare and quantify
the nature of absorption-selected galaxies.

Furthermore, we have extracted absorption parameters from the quasar
spectra as well as from Voigt profiles fits to the absorption systems.
We have measured the optical depth weighted mean {\MgII} $\lambda2796$
absorption redshift, the rest-frame equivalent width $W_r(2796$), the
doublet ratio, the number of clouds, the Voigt profile fitted system
column density, the {\MgII} optical depth, the AOD derived column
density, the mean velocity, the velocity spread, and the velocity
asymmetry.

In order to explore possible connections between the {\MgII}
absorption properties and the galaxy morphological properties, we have
performed non-parametric Spearman and Kendall rank correlation tests.
Our mains results can be summarized as follows:

\begin{enumerate}
\item~\MgII~host galaxies appear to be similar to those at low
redshift and have a wide range of morphologies and colors. With $27$
late-type galaxies and $13$ early-type galaxies, their populations are
consistent with the distribution of galaxy morphologies found in the
field environment. They have a mean disk scale length of 3.8~kpc and a
mean bulge scale length 2.5~kpc, which are comparable those of the
Milky Way.  The disk scale lengths and bulge effective radii
distributions of the sample are similar to those of field galaxies
obtained from the DEEP survey, at similar redshifts.  The mean
$W_r(2796)$ for the early-type galaxies is 1.0~{\AA} and slightly
lower for late-type galaxies at 0.7~{\AA}, where both types have a
large spread in equivalent width, although late-type galaxies dominate
for absorption systems with $W_r(2796)<0.3$~{\AA} and tend to also be
at $D>40$~kpc.

\item
We find a $3.2\sigma$ anti-correlation between $D$ and {\MgII}
equivalent width. There is large a scatter in the distribution,
suggesting that $D$ is not the only physical parameter affecting the
distribution and quantity of halo gas. There are no other absorption
properties that scale with $D$ above $3\sigma$. However, {\it all\/}
our other measures of the gas absorption strength show decreasing
trends with $D$.  Taken together, the anti-correlation trends show a
consistent signature that, on average, {\MgII} absorbing gas becomes
optically thinner with increasing projected galactocentric distance
from the host galaxy and is likely the dominant behavior of the halo
gas.

\item
We find only weakly significant trends ($2-2.2\sigma$) between color
and the absorption properties, suggesting that larger absorption
quantities tend to be associated with the bluer galaxies. We
do not reproduce the $B-K$ and $W_r(2796)$ correlation
of \citet{zibetti07}. Our results are consistent with \citet{chen10}
and we conclude that our samples probe a lower equivalent width range
then \citeauthor{zibetti07}, and therefore are likely probing
different mechanisms producing {\MgII} absorption. We find less than a
$1\sigma$ connection between $M_B$ and the absorption properties,
implying that for our luminosity and equivalent width range the
{\MgII} absorption is not strongly dependent on galaxy luminosity.

\item
We find no correlation between $\theta$ and the absorption properties
(the highest significance is 1.4$\sigma$); the absorption properties
are consistent with a random distribution as a function of position
angle. 

\item By accounting for the decreasing gas column density (and its
scatter) with increasing $D$, the correlation with $i/D$ and
$W_r(2796)$ increases to $4.3\sigma$ significance level. Also,
following normalization of $i$ by $D$, we find that $N_a$, $\tau$,
$N_{cl}$, $N_{vp}$ and $W_r(2803)$ are all correlated with $i/D$
greater than $3.2\sigma$ level of significance.  Overall, these
results suggest that the {\MgII} gas has co-planer geometry, but is
not necessarily disk-like, that is coupled to the galaxy
inclination. This results do not support {\MgII} absorption produce by
star-burst driven winds. These results are consistent with the models
of \citet{stewart11}.

\item
We do not find any other correlations above $2.0\sigma$ between the
remaining galaxy properties, such as $r_b$, $r_h$, $r_e$, $B/T$, $n$
and the absorption properties. Thus, the galaxy optical size and shape
does not appear to be an important factor in determining the amount of
gas within a galaxy halo.

\end{enumerate}

We find several interesting connections between the galaxy
morphological properties and the {\MgII} absorbing gas.  Although
recent evidence suggests that high equivalent width systems are
produced predominantly by winds in star-forming galaxies, the results
of this paper do not support such an explanation for weaker systems
($W_r(2796)\la 1$\,\AA). The correlation between the inclination of
the galaxy disk and the halo gas absorption strength is suggests that
the the halo gas resides in a co-planer distribution. It is plausible
that the absorbing gas arises from tidal streams, satellites,
filaments, etc. which tend to have more-or-less co-planer
distributions. This correlation could not be explained in the wind
scenario.

One of the few ways to differentiate between winds and other sources
replenishing the halo gas is by studying multi-phase gas sensitive to
different density and temperatures.  These questions express the
necessity of UV spectrographs like COS and STIS.


\section*{Acknowledgments}

We thank Simon Mutch for his contributions to the manuscript.  We also
thank Nicolas Bouch\'{e} and Simon Lilly for careful reading
manuscript and providing comments. We thank the anonymous referee for
providing insightful comments and improving the paper. M.T.M thanks
the Australian Research Council for a QEII Research Fellowship
(DP0877998).  CWC and JLE were partially supported by the National
Science Foundation under Grant Number AST-0708210.  This work is based
in part on observations made with the NASA/ESA {\it Hubble Space
Telescope}, or obtained from the data archive at the Space Telescope
Institute (STScI), which is a collaboration between STScI/NASA, the
Space Telescope European Coordinating Facility (ST-ECF/ESA) and the
Canadian Astronomy Data Centre (CADC/NRC/CSA).  Other observations
were obtained with the European Southern Observatory (ESO) Very Large
Telescope at the Paranal Observatories and with the W.M. Keck
Observatory (some of which were generously provided by Jason
X. Prochaska and by Wallace L. W. Sargent and Michael Rauch), which is
operated as a scientific partnership among the California Institute of
Technology, the University of California and the National Aeronautics
and Space Administration. Keck Observatory was made possible by the
generous financial support of the W.M. Keck Foundation.


\newpage
\appendix
\section{APPENDIX A: Supporting Information}

Figure A1 is the complete version of Figure~\ref{fig:mos}. It will appear in the
online version of this paper, not in the printed version.

\begin{figure*}
\includegraphics[angle=0,scale=0.43]{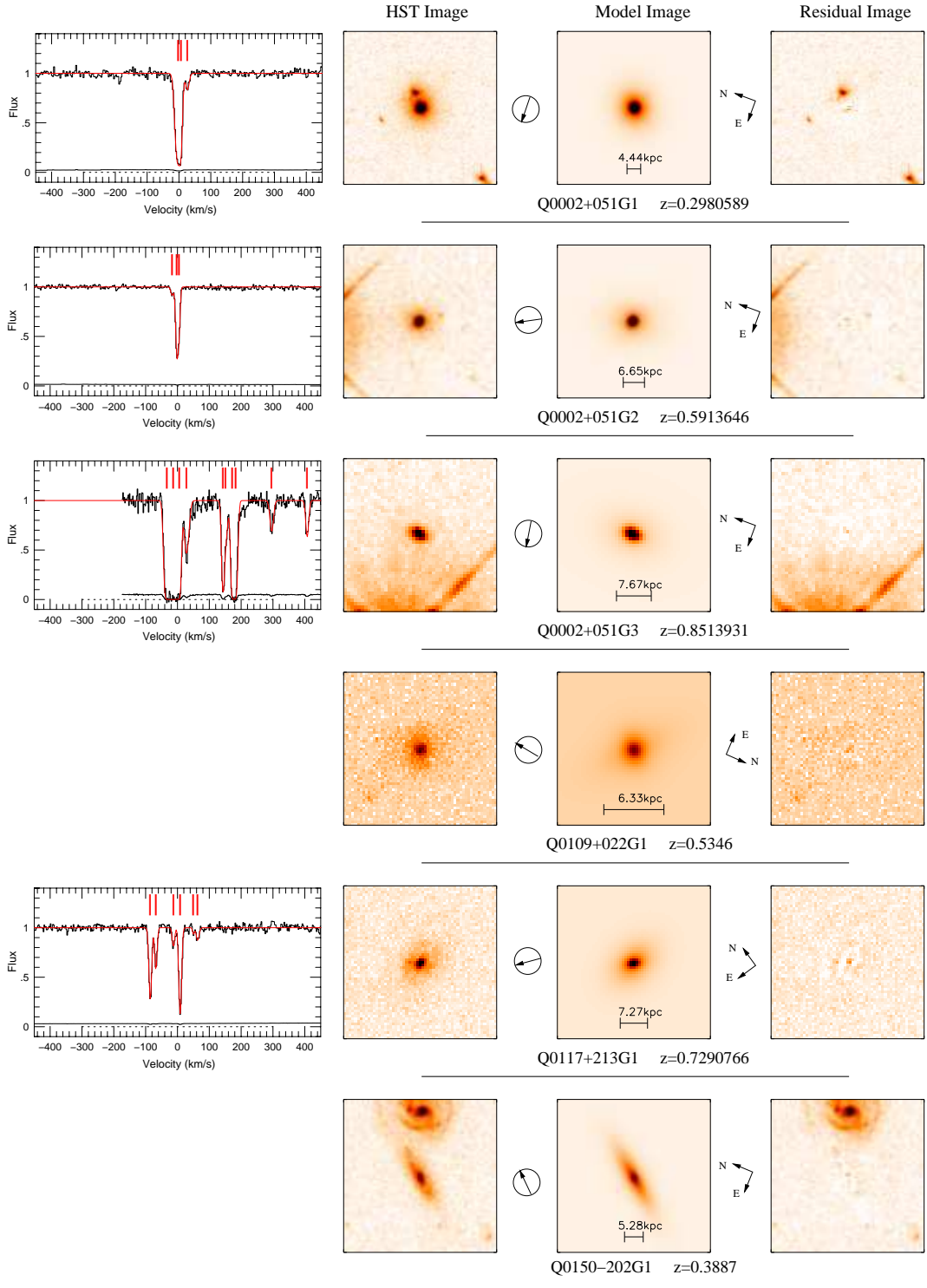}
\caption{--- (far-left) The HIRES/Keck or UVES/VLT quasar spectra of
the {{\rm Mg}\kern 0.1em{\sc ii}~$\lambda 2796$} absorption feature
are shown alongside the associated absorbing galaxy on the right. The
{\MgII} $\lambda 2796$ optical depth weight mean absorption redshift
is the zeropoint of the velocity scale. The tick marks indicate the
number of Voigt profile components and the red curve indicates the fit
to the data.  We do not have HIRES or UVES data available for six
galaxies. (left) WFPC-2/{\it HST} images of galaxies selected by
{\MgII} absorption. The images are 10 times larger than the
$1.5\sigma$ isophotal area.  --- (center) The GIM2D models of the
galaxies, which provide quantified morphological parameters. A scale
of one arcsecond is indicated on each image along with the physical
scale computed at the {\MgII} absorption redshift.  --- (right) The
residual images from the models, showing quality of the fit and the
underlying structure and morphological perturbations of the
galaxies. The encircled arrow provides the direction to the quasar
(galaxy--quasar orientation). The cardinal directions are also shown
and the quasar name and redshift of {\MgII} absorption is provided
under each set of galaxy WFPC-2, model and residual image.}
\end{figure*}

\addtocounter{figure}{-1}
\begin{figure*}
\includegraphics[angle=0,scale=0.43]{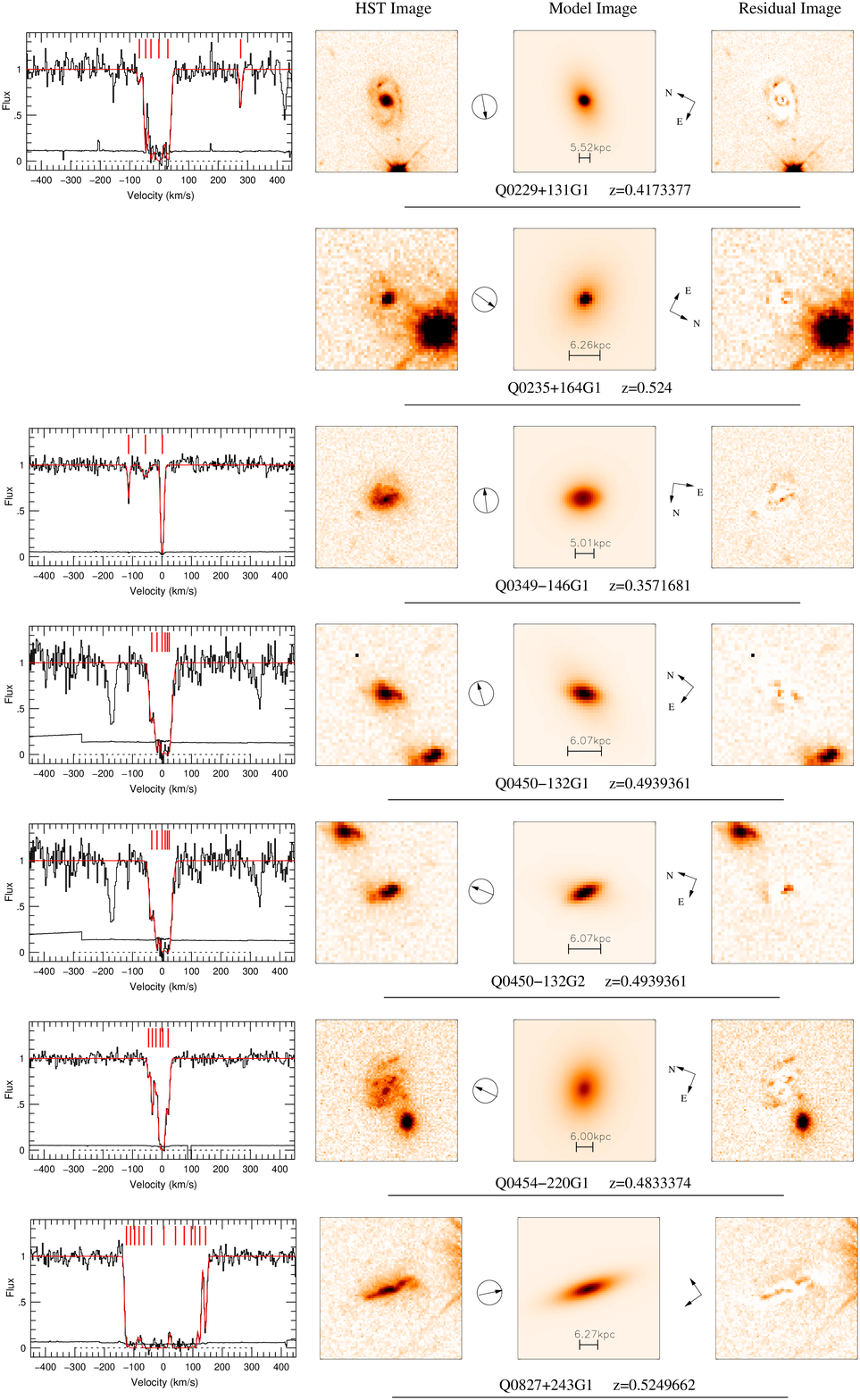}
\caption{--- continued}
\end{figure*}
\addtocounter{figure}{-1}
\begin{figure*}
\includegraphics[angle=0,scale=0.43]{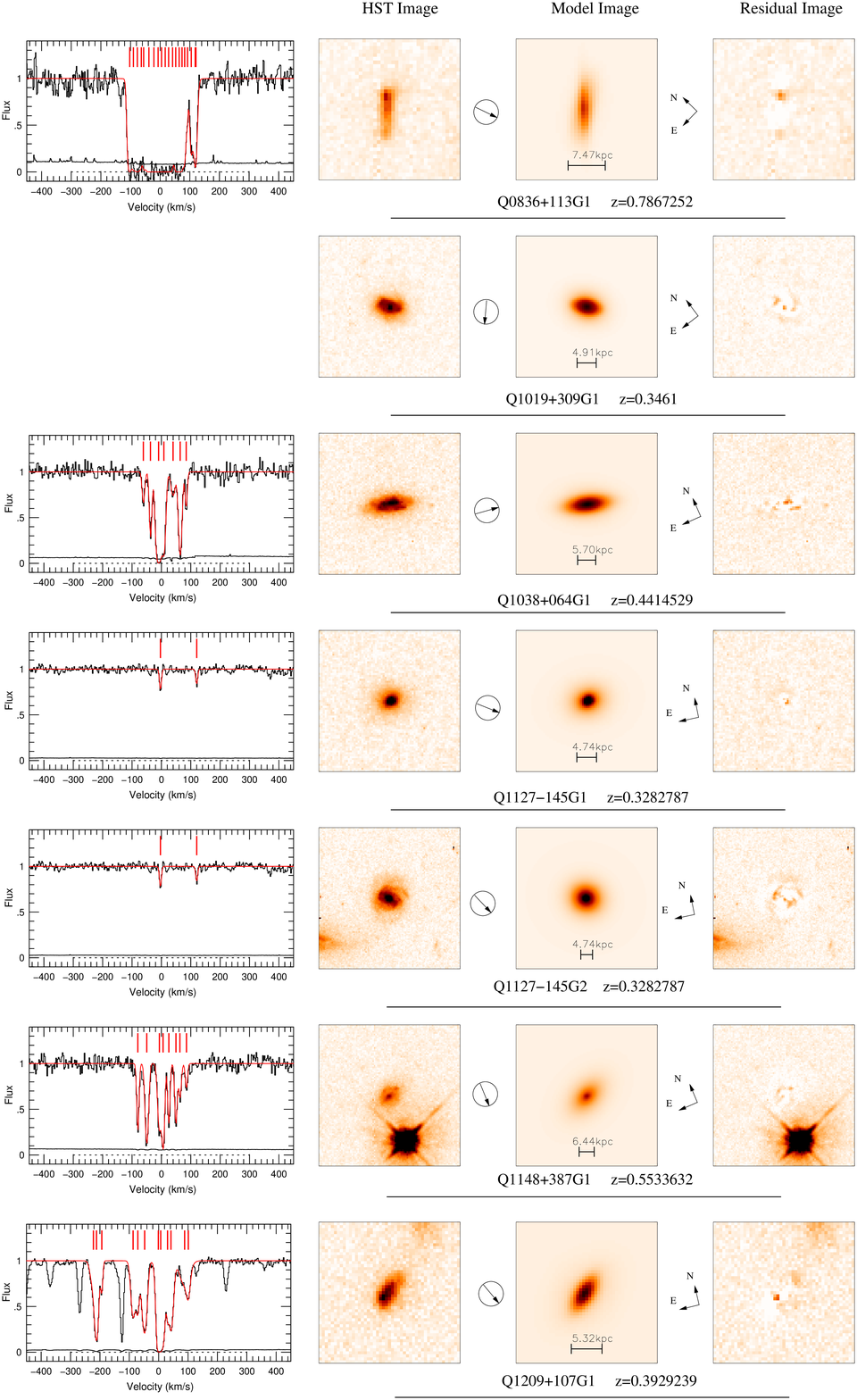}
\caption{--- continued}
\end{figure*}
\addtocounter{figure}{-1}
\begin{figure*}
\includegraphics[angle=0,scale=0.43]{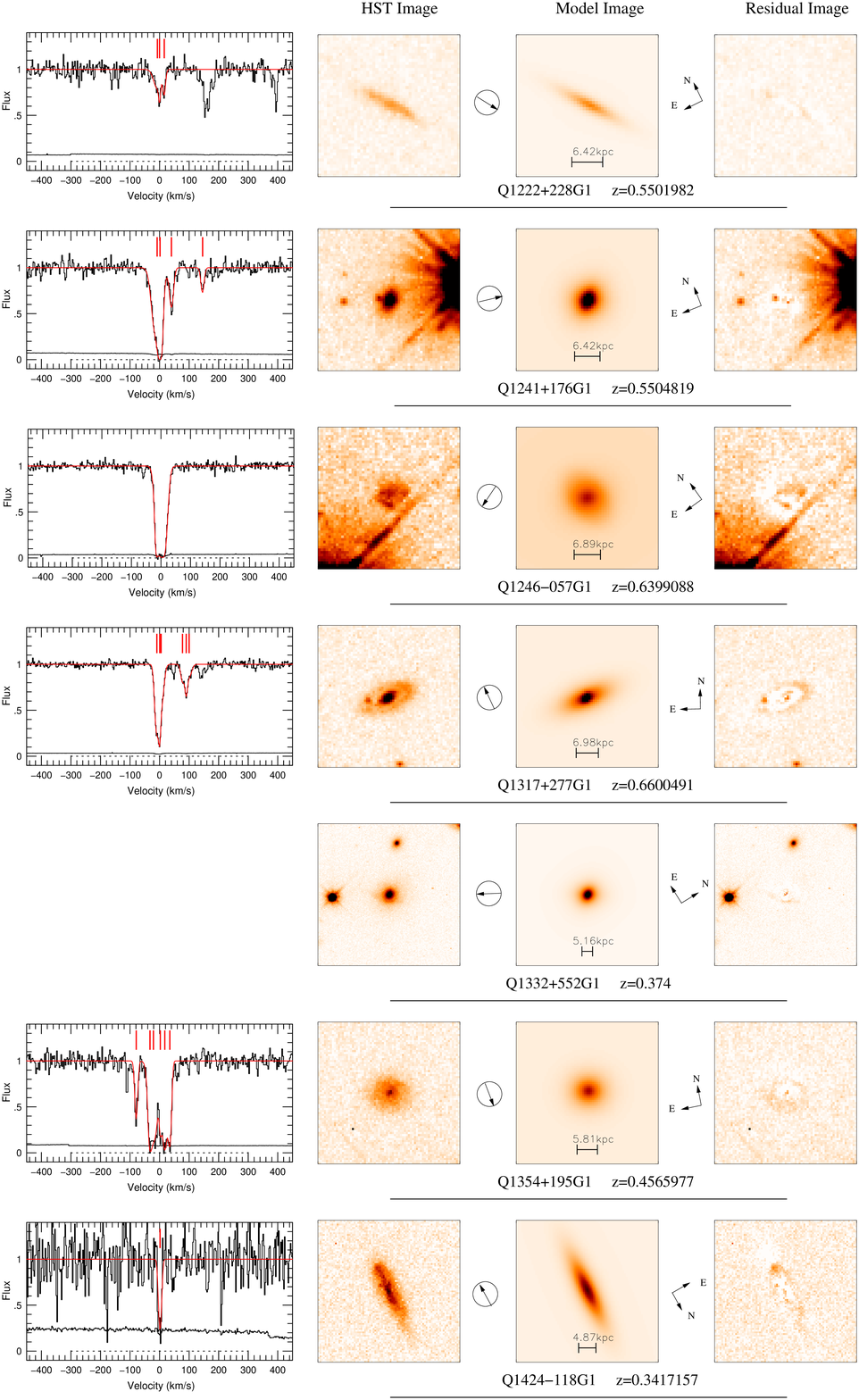}
\caption{--- continued}
\end{figure*}
\addtocounter{figure}{-1}
\begin{figure*}
\includegraphics[angle=0,scale=0.43]{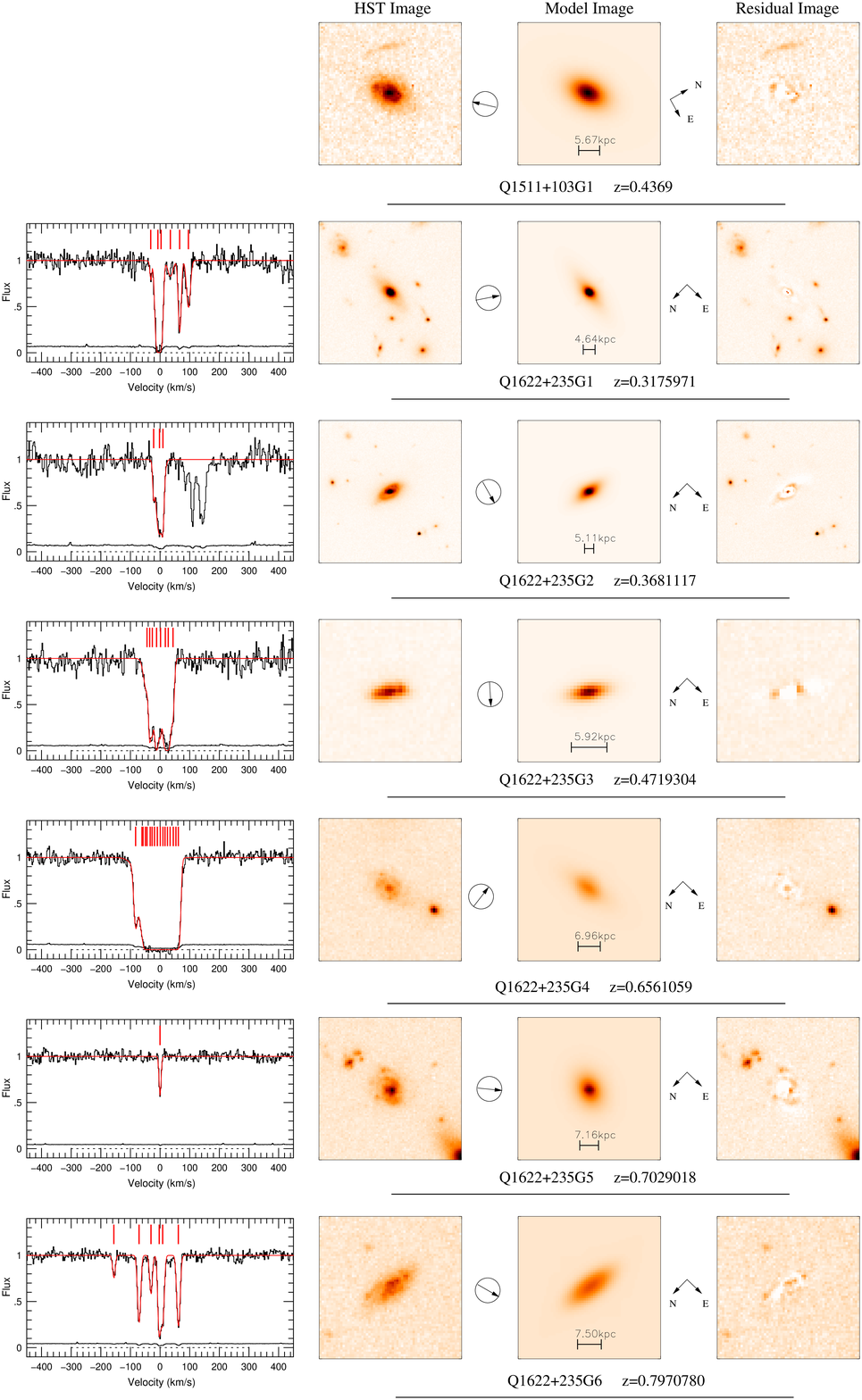}
\caption{--- continued}
\end{figure*}
\addtocounter{figure}{-1}
\begin{figure*}
\includegraphics[angle=0,scale=0.43]{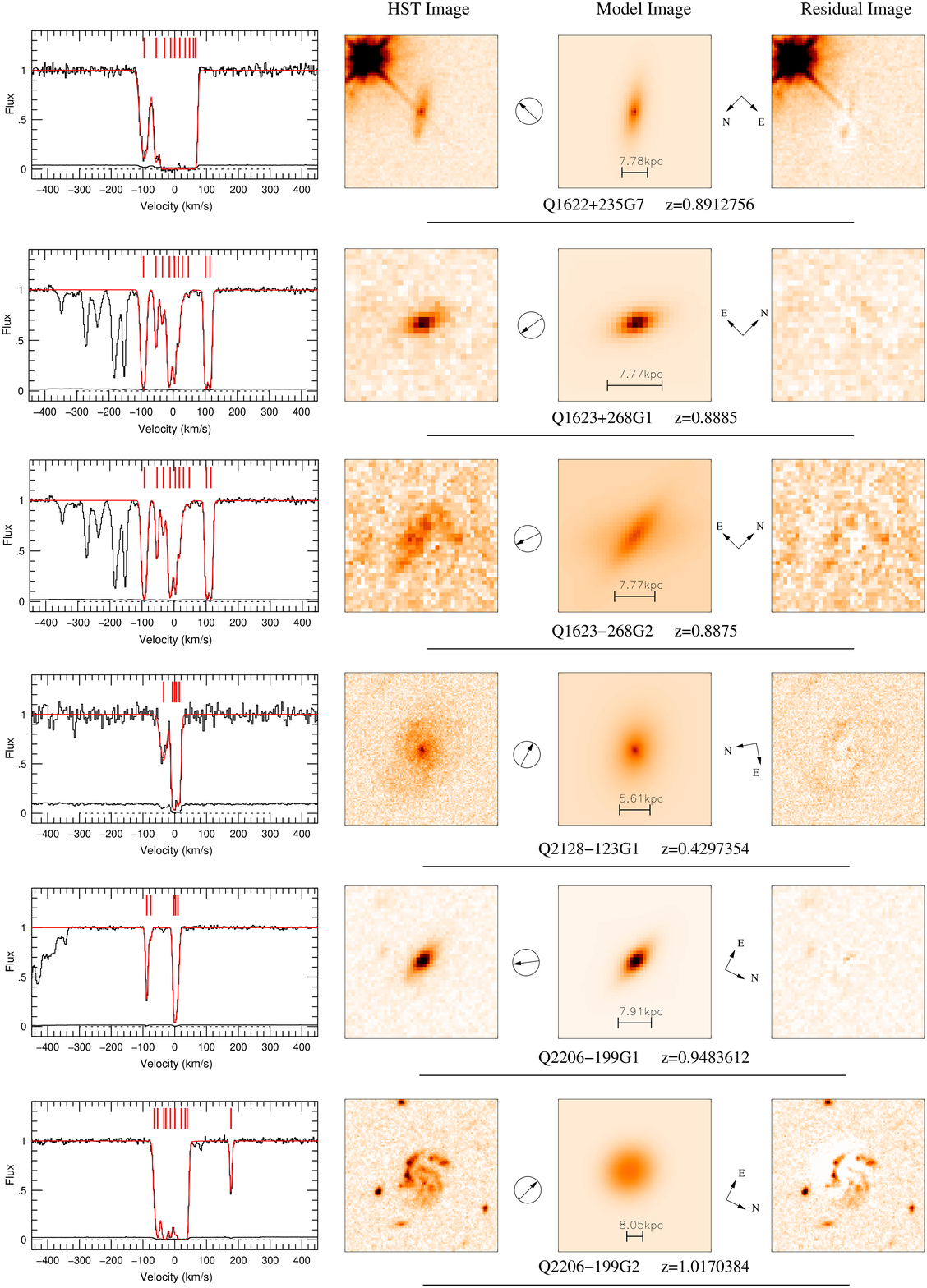}
\caption{--- continued}
\end{figure*}

\label{lastpage}

\end{document}